\documentclass[journal,onecolumn]{IEEEtran}

\usepackage{tikz}
\usetikzlibrary{arrows}

\usepackage{setspace}
\usepackage{graphicx}
\usepackage{amsmath}
\usepackage{amssymb}
\usepackage{amsthm}
\usepackage{algorithmicx}
\usepackage[Algorithm,ruled]{algorithm}
   \usepackage{algpseudocode}
\floatstyle{boxed}
\usepackage{verbatim}
\restylefloat{figure}
   
\usepackage[latin1]{inputenc}
\usepackage{amsfonts}
\usepackage{fullpage}
\usepackage{hyperref}
\usepackage{cleveref}
\usepackage{cleveref}
\usepackage{breqn}
\usepackage{enumerate}
\usepackage{enumitem}
\usepackage{soul}

\setkeys{breqn}{breakdepth={1}}

\theoremstyle{plain}
\newtheorem{theorem}{Theorem}

\newtheorem{lemma}[theorem]{Lemma}

\usepackage[backend=biber]{biblatex}
\theoremstyle{definition}

\newtheorem{definition}[theorem]{Definition}
\newtheorem{example}{Example}

\crefname{equation}{inequality}{inequalities}
\Crefname{equation}{Inequality}{Inequalities}

\title{Efficient Decoding of Insertion and Deletion Errors for Helberg Codes}
\author{
    \IEEEauthorblockN{Anthony Segrest\IEEEauthorrefmark{1}, Hieu D. Nguyen\IEEEauthorrefmark{2}} \\
    Rowan University \\
    \IEEEauthorblockA{\IEEEauthorrefmark{1}segres62@students.rowan.edu},
    \IEEEauthorblockA{\IEEEauthorrefmark{2}nguyen@rowan.edu} \\
    8-26-2025
}

\date{Original date: 10-18-2019, Last update: 1-27-2025}

\begin{document}

\maketitle

\begin{abstract}
We present the first known efficient decoding algorithm for correcting multiple insertion-deletion errors in Helberg codes and their non-binary generalizations, extending a known algorithm for correcting multiple deletion errors.
\end{abstract}

\begin{IEEEkeywords}
Insertion deletion error-correcting codes, Helberg codes.
\end{IEEEkeywords}

\section{Introduction}


Helberg codes 
are binary error-correcting codes, designed to correct up to a combination of $k$ insertion-deletion errors. First introduced by Helberg and Ferreira in \cite{helberg_code_paper} and further investigated by Abdel-Ghaffar et al \cite{helberg_generalization} among others, Helberg codes are number-theoretic codes generalizing Varshamov-Tenengolts (VT) codes, constructed by Varshamov and Tenengo'lts \cite{VT} to correct a single asymmetrical error and later proved by Levenshtein \cite{L} to be capable of also correcting a single insertion or deletion error (indel for short). In the same paper, Levenshtein also showed that if a codebook entirely of strings of length $n$ can correct $d$ deletions, it can correct any $d$ combination of insertion and deletion errors (collectively known as indels). Levenshtein also found a linear-time algorithm for VT codes to correct a single deletion. 

VT codes have asymptotically optimal redundancy, but until recently, asymptotically optimal redundancy was not achieved for binary codes that correct multiple deletions. Helberg codes appear to have $\Omega(n)$ redundancy as mentioned in \cite{8022906}, though the asymptotic rate of Helberg codes is unknown. The cardinality of Helberg codes can be calculated by formula (albeit an inefficient one), derived by Bibak and Milenkovic \cite{weight_enumerators}, who also gave an upper bound; however, the explicit rate of Helberg codes remains unknown.

Recently, a binary code with $O(d^2 \log d \log n)$ redundancy was constructed in \cite{8022906} by Brakensiek, Guruswami, and Zbarsky. Following this, binary codes with redundancy $O(d \log n)$ were created by Cheng et al. \cite{8555106}. More recently, a binary code with redundancy $O(d \log n)$ (in particular, $8 d\log n + o(\log n)$) was created by Sima and Bruck - this code is the first to achieve asymptotically optimal redundancy for all $n$ and $k$ \cite{sima_asymptotically_optimal_redundancy}. Shortly thereafter, Sima, Gabrys, and Bruck  \cite{systematic_code} created a systematic code with redundancy $4 d\log n + o(\log n)$ and later created asymptotically optimal codes for non-binary alphabets \cite{non_binary_alphabet_optimal}.

Helberg codes have been generalized to non-binary alphabets by Le and Nguyen \cite{deletion_algo_paper}, who also gave in the same paper a polynomial-time decoding algorithm when the codeword has only suffered deletion errors (technically, the algorithm there includes a section that runs in $O(q^d)$ time, where $q$ is the alphabet size and $d$ is the number of deletions that can be corrected, but this can be easily adapted to run in polynomial time in $n$ and $d$). Also, in \cite{helberg_abstractions}, an abstraction of Helberg codes is used to create families of new deletion-correcting codes.

In this paper we present a polynomial-time (in terms of $n$, the length of the codeword) decoding algorithm to correct multiple insertion-deletion errors for both Helberg codes and their non-binary generalization \cite{deletion_algo_paper}. It is the first efficient algorithm for decoding Helberg codes, and uses the decoding algorithm for correcting only deletions from \cite{deletion_algo_paper} as a sub-step (and also shares some similar ideas such as decoding from right-to-left to recover the correct codeword one bit (or a group of bits) at a time.

To explain our algorithm, we begin with formally defining Helberg codes and their non-binary generalization. Let $C$ be a code consisting of strings of length $n$, $d$ be the maximum number of indels that $C$ is capable of correcting, and $q$ be the size of the alphabet. The codewords in $C$ are denoted by $\mathbf{x}=(x_1,...,x_n)\in A^n$, where $A=\{0,1,...,q-1\}$ specifies our $q$-ary alphabet.  We employ the notation $x_i$ to denote the $i$-th bit of $\mathbf{x}$. Also, we set $p = q - 1$ to be the maximum possible value of any bit in $\mathbf{x}$.

The generalization of Helberg codes as described in \cite{deletion_algo_paper} can now be constructed as follows: define a sequence of weights $W(q,d)=\{w_1(q,d),w_2(q,d),...\}$ where we initialize $w_i(q,d)=0$ if $i\leq 0$.  Then for $i\geq 1$, we define $w_i(q,d)$ recursively by
\begin{equation}
w_i(q,d)=1+p\sum_{j=1}^d w_{i-j}(q,d).
\end{equation}
Where the value of $q$ and $d$ is clear, we shall write $w_i$ in place of $w_i(q, d)$. Next, we define the moment of a codeword $\mathbf{x}$, written as $M(\mathbf{x})$, by the formula
\begin{equation}
    M(\mathbf{x}) = \sum_{i=1}^{n}x_iw_i
\end{equation}

 We are now ready to define our desired codes:

\begin{definition} \label{de:generalized-helberg}
Let $n$, $m$ and $r$ be fixed integers satisfying $m\geq w_{n+1}$ and $0\leq r < m$.  We define the code $C_H(n,d,r,q,m)$ to be the set of codewords of length $n$ whose moments have residue $r$ modulo $m$, i.e.,
\[
C_H(n,d,r,q,m)=\{\mathbf{x}\in A^n: M(\mathbf{x}) \equiv r \ \mathrm{mod} \ m \}.
\] 

\begin{theorem}[\cite{deletion_algo_paper}]
These codes have been proven capable of correcting up to $d$ insertion and deletion errors.
\end{theorem}

\end{definition}
\noindent For convenience, we set $m = w_{n+1}$ for the rest of this paper and use the abbreviated notation $C_H(n,d,r,q) \equiv C_H(n,d,r,q,w_{n+1})$ for convenience. In the case of a binary alphabet where $q=2$, the codes $C_H(n,d,r,2,m)$ are referred to as Helberg codes.

Our main result is a new efficient decoding algorithm for $C_H(n, d, r, q)$, called the Indel-Correcting Algorithm, to recover any codeword $\mathbf{x} \in C_H(n, d, r, q)$ from its corruption, denoted by $\mathbf{y}$, where at most $d$ indels occurred.  We describe this algorithm in detail in Section  \ref{Algorithm} and also present an algorithm, called the Moment Algorithm, to determine the exact moment of a corrupted codeword $\mathbf{y}$ which is used at the start of the Indel-Correcting Algorithm.

From a high-level perspective, the Indel-Correcting Algorithm works by decoding bits of $\mathbf{x}$ from right to left, and looping until the full word has been decoded. Inside each loop are two steps, referred to as Step 1 and Step 2:
\begin{itemize}
    \item Step 1: We use the fact that for each possible value of the rightmost unknown bit, there is a range of possible values of $M(\mathbf{x})$. If $M(\mathbf{x})$ falls into only one of these ranges, we learn the value of the rightmost unknown bit immediately. If we are successful, we restart the loop, going back to Step 1.
    \item Step 2: If Step 1 cannot decode the rightmost unknown bit, the loop moves to Step 2, which decodes $d$ bits or the entire codeword. In Step 2, we split the possibilities into two cases - one of them will contain a long string of zeros, and the other will contain a long string of $p$'s.. Then, we test which case will be more efficient to explore completely by computing, roughly speaking, the number of bits at the right end of $\mathbf{y}$ that are ``used up'' in each case by the requirement that the longest common subsequence between $\mathbf{x}$ and $\mathbf{y}$ is sufficiently large. The more bits of $\mathbf{y}$ that are used up, the faster it is to explore the case. So, we explore the case with more bits used up, and either find the solution (decoding the entire string), or rule it out (meaning the other case must be correct, which decodes $d$ bits). Due to a bound on the maximum number bits of $\mathbf{y}$ are used up between the two cases, the time complexity of this step (and the whole algorithm) is polynomial, with a constant exponent. In this step, we sometimes use the Deletions Algorithm from \cite{deletion_algo_paper} as a substep. 

\end{itemize}

In Section III, we provide a proof of our Indel-Correcting Algorithm and then present examples to demonstrate explicitly the steps involved in the algorithm in Section IV. Lemmas describing certain properties of strings and their longest common subsequences, which are needed in the proof of Indel-Correcting Algorithm, are justified in the Appendix. A Python implementation of our Indel-Correcting Algorithm is available on Github \cite{githubimplementation}.




\section{Indel-Correcting Algorithm} \label{Algorithm}
\newcommand{\ssep}{\mid}

We begin by defining notation needed to explain our algorithm.

\begin{enumerate}
\item For a string $s$, $len(s)$ is the number of characters in $s$.
\item For an integer $j$, $[j]$ is the string of length 1 containing $j$.
\item For a string $s$ and integers $j_1$ and $j_2$, with $0 \leq j_1 \leq j_2$, $s[j_1:j_2] := s_{j_1}s_{j_1+1}...s_{j_2}$.
\item For a string $s$ and integers $j_1$ and $j_2$, with $j_1 > j_2$, $s[j_1:j_2]$ is $\Lambda$, the empty string. 
\item For integers $a \geq 0$, and $0 <= b <= p$ (where $p$ is the largest number in the alphabet), $[b] * a := bb...bb$ (a string of $b$'s, of length $a$.)
\item For two strings $s$ and $r$, $s + r := s_1s_2...s_{len(s)}r_1r_2...r_{len(r)}$.
\item For integers $j_1$ and $j_2$, and function $f(i)$, where $j_1 > j_2$, $\sum_{i=j_1}^{j_2} f(i) = 0$.
\item For two strings $s_1$ and $s_2$, $lcs(s_1, s_2)$ is the length of the longest common subsequence between $s_1$ and $s_2$.
\item Define $a = \lfloor\frac{d + len(\mathbf{y}) - n}{2}\rfloor\ $
and $b = \lfloor\frac{d - len(\mathbf{y}) + n}{2}\rfloor$.
\end{enumerate}

The values $a$ and $b$ give upper bounds on the number of insertion and deletion errors respectively, that occurred in a given codeword (see \Cref{lem:a_b_bounds}). $a$ and $b$ both can never be negative, because we assume $\mathbf{y}$ has had at most $d$ indel errors. We either have $a + b = d - 1$ or $a + b = d$. However, for the Indel-Correcting Algorithm, we shall assume $a + b = d$, which can be achieved by further corrupting the codeword with one more deletion and simplifies the proof of the algorithm. Observe that $len(\mathbf{y}) = n + a - b$.

If $d = 1$ (single insertion or deletion), a simple brute force approach of checking each possible insertion or deletion should be used. Otherwise, our algorithm first determines $x_n$, then $x_{n - 1}$, etc. Therefore, we must prove that, for any $1 \leq n' \leq n$, if we have correctly determined the values of $x_i$ for all $n' < i \leq n$, then we can also correctly determine $x_{n'}$.

A priori, given $\mathbf{x} \in C_H(n,d,r,q) $, we only know the residue of its moment $M(\mathbf{x})$ ($\textrm{mod} \ w_{n+1}$); however, $M(\mathbf{x})$ can be determined exactly using our Moment Algorithm (\Cref{thm:moment_thm_p1,thm:moment_thm_p2}).  The Indel-Correcting Algorithm requires knowing the moment of $\mathbf{x}$ and its corresponding partial moments, i.e., the moment of the first $n'$ bits of $\mathbf{x}$: 
\begin{equation}
m' := \sum_{i=1}^{n'} x_im_i = M(\mathbf{x}) - \sum_{i=n' + 1}^{n} x_iw_i
\end{equation}

We now describe the Indel-Correcting Algorithm.
\subsection{\bf Indel-Correcting Algorithm}

\noindent\hspace*{-1.5em}{\bf Preliminaries:}

\noindent Perform the following if necessary:

\begin{enumerate}[label=P{\arabic*}:]
    \item \label{algo:a_b_d_fix}
If $a + b < d$, then make one arbitrary deletion, and instead perform the algorithm on this new codeword. Therefore, we now have $a + b = d$.

    \item If $d = 1$, simply check each possible insertion or deletion with brute force, instead of doing the algorithm below.

    \item \label{algo:moment_algorithm_usage}
Use the Moment Algorithm (\ref{algo:moment_algorithm_definition}) to find the value of $M(\mathbf{x})$.

    \item Initialize $n' = n$.

\end{enumerate}


\noindent\hspace*{-1.5em}{\bf While $n' > 0$:}

\begin{enumerate}[label=\textbf{Step {\arabic*}}]
\item 
\label{step:1} Let $H = \{0 \leq g \leq p \ssep gw_{n'} \leq m' \leq gw_{n'} + p\sum_{i=1}^{n' - 1} w_i\}$. Let $g = min(H)$. If $|H| = 1$, then $x_{n'} = g$ (proven in \Cref{lem:inrange}). Decrement $n'$, and repeat this step.

\item 
\label{step:2} Otherwise, there are only two possible cases (proven in \Cref{thm:two_cases}):

\begin{enumerate}[label={\alph*})]
    \item $x_{n'} = g + 1$ and $x_{i} = 0$ for all $n' - d < i < n'$. (Case 1)
    \item $x_{n'} = g$ and $ x_{i} = p$ for all $n' - d < i < n'$. (Case 2)
\end{enumerate}

We define $\mathbf{x}^{(1)} = [0] * (d - 1) + [g + 1] + x[n' + 1 : n]$ and $\mathbf{x}^{(2)} = [p] * (d - 1) + [g] + x[n' + 1 : n]$ corresponding to Case 1 and 2, respectively.

Now:  

\begin{enumerate}[label={\roman*})]

\item \label{def:v_definition} For $i \in \{1, 2\}$, compute the lowest nonnegative integer $v_i$ such that $lcs(\mathbf{x}^{(i)}, \mathbf{y}[len(\mathbf{y}) - v_i + 1 : len(\mathbf{y})]) \geq n - n' + d - b$, or if no such integer exists, we let $v_i = \infty$. Let $t$ be the index such that $v_t= \max(v_1, v_2)$ (if $v_1 = v_2$, then either value of $t$ can be chosen). 
The value of $v_t$ determines what we do next.

\item \label{algo:step_2_best_case} If $v_t \geq n - n' + 2a + 1$, then we have proven this case false, and should go to \textbf{Step 2 \ref{larger_v_case_proven_false}} (proven in \Cref{thm:v_upper_bound}).



\item \label{algo:step_2_middle_case} If $v_t = n - n' + 2a$, let $m'' = M(\mathbf{x}) - \sum_{i=n' - d + 1}^{n}x_{i-n'+d}^{(t)}w_i$. Use the Deletions Algorithm for the codebook $C_H(n' - d, d, m'', q)$ to decode $\mathbf{y'} \equiv \mathbf{y}[1 : len(\mathbf{y}) - v_t - b]$. Call the result of decoding $\mathbf{x'}$.
    
    If $M(\mathbf{x'} + \mathbf{x}^{(t)}) \neq M(\mathbf{x})$, or $lcs(\mathbf{x'} + \mathbf{x}^{(t)}, \mathbf{y}) < n - b$ then we have proven this case false and should go to \textbf{Step 2 \ref{larger_v_case_proven_false}}. Otherwise, we know $\mathbf{x} =\mathbf{x'} + \mathbf{x}^{(t)}$, and the algorithm is complete (proof in \Cref{thm:v_is_n_n_prime_2a,lem:check_if_correct}).

\item \label{algo:step_2_worst_case} If $v_t = n - n' + 2a - 1$, let $m'' = M(\mathbf{x}) - \sum_{i=n' - d + 1}^{n}x_{i-n'+d}^{(t)}w_i$.

\noindent If $len(\mathbf{y}) - v_t - b > 0$:
\begin{itemize}
\item
For each $1 \leq j \leq len(\mathbf{y}) - v_t - b$, use the Deletions Algorithm for the codebook $C_H(n' - d, d, m'', q)$ to decode $\mathbf{y'} \equiv \mathbf{y}[1:j-1] + \mathbf{y}[j + 1 : len(\mathbf{y}) - v_t - b]$. Call the result of decoding $\mathbf{x'}$.

For each $j$, check if $M(\mathbf{x'} + \mathbf{x}^{(t)}) = M(\mathbf{x})$ and if $lcs(\mathbf{x'} + \mathbf{x}^{(t)}, \mathbf{y}) \geq n - b$. If this is true for some $\mathbf{x'}$, then we have $\mathbf{x} = \mathbf{x'} + \mathbf{x}^{(t)}$ and the algorithm is complete. If no $\mathbf{x'}$, then we have proven this case false and should go to \textbf{Step 2 \ref{larger_v_case_proven_false}}.

\end{itemize}
\noindent If $len(\mathbf{y}) - v_t - b \leq 0$:
\begin{itemize}
\item
Use the Deletions Algorithm for the codebook $C_H(n' -d, d, m'', q)$ to decode $\mathbf{y'} \equiv \Lambda$ (the empty string). Call the result $\mathbf{x'}$.

If $M(\mathbf{x'} + \mathbf{x}^{(t)}) \neq M(\mathbf{x})$, or $lcs(\mathbf{x'} + \mathbf{x}^{(t)}, \mathbf{y}) < n - b$ then we have proven this case false and should go to \textbf{Step 2 \ref{larger_v_case_proven_false}}. Otherwise, we have found $\mathbf{x} = \mathbf{x'} + \mathbf{x}^{(t)}$, so the algorithm is complete (proof in \Cref{thm:v_is_n_n_prime_2a_minus_1,lem:check_if_correct}).

\end{itemize}

\item \label{larger_v_case_proven_false} If we have reached this step, we know that the case with the smaller value of $v$ is correct (remaining proof in \Cref{thm:v_lower_bound}). This means we have determined the values of $x_j$ for all $n' - d < j \leq n'$, and we go back to step 1, setting $n'$ to $n' - d$.

\end{enumerate}

\end{enumerate}

Note: We sometimes give invalid inputs to the Deletions Algorithm in parts iii and iv of Step 2 - in other words, we give the Deletions Algorithm a corrupted codeword $\mathbf{y'}$ and Helberg codebook of shorter length, such that there is no $\mathbf{x'}$ in the codebook that could become $\mathbf{y'}$ with an allowed number of deletions (the allowed number being determined by the parameters of the codebook). Our algorithm does not depend on the behavior of the Deletions Algorithm in these cases, except for the fact that the Deletions Algorithm will still output a string of the correct length.

\begin{algorithm}[H]
 \floatname{algorithm}{Indel Algorithm}
 \caption{}
\begin{algorithmic}[1]
\Function{Decode}{$\mathbf{y}$, $n$, $q$, $d$, $r$}
\If {$d == 1$}
    \State \Return DecodeBruteForce($\mathbf{y}$, $n$, $q$, $r$)
\EndIf
\State $a = \lfloor\frac{d + len(\mathbf{y}) - n}{2}\rfloor$
\State $b = \lfloor\frac{d - len(\mathbf{y}) + n}{2}\rfloor$
\If {$a + b < d$}
    \State \Return Decode($\mathbf{y}[2:len(\mathbf{y})]$, $n$, $q$, $d$, $r$)
\EndIf
\State $M(\mathbf{x}) = $ FindOriginalMoment($\mathbf{y}$, $n$ $d$, $r$) \Comment{This is the Moment Algorithm below}
\State $\mathbf{x} = [0] * n$ \Comment{$\mathbf{x}$ here represents what we have found so far, with the $0$'s being placeholders}
\State $n' = n$
\While {$n' > 0$}
    \State $m' = M(\mathbf{x}) - \sum_{i=n' + 1}^{n}x_iw_i$
    \State $H = \{0 \leq g \leq p \ssep gw_{n'} \leq m' \leq gw_{n'} + p\sum_{i=1}^{n'-1}w_i\}$
    \State $g = min(H)$
    \If {$|H| == 1$}
        \State $x_{n'} = g$
        \State $n' = n' - 1$
    \Else
        \State lower\_g\_correct, possible\_full\_solution = DoStepTwo($\mathbf{x}$, $\mathbf{y}$, $d$, $q$, $m$, $n$, $n'$, $a$, $g$) \Comment{Explained in page below}
        \If {possible\_full\_solution $\neq \Lambda$}
            \State \Return possible\_full\_solution
        \ElsIf {lower\_g\_correct}
            \State $\mathbf{x}[n' - d + 1:n'] = [0] * (d - 1) + [g + 1]$
        \Else
            \State $\mathbf{x}[n' - d + 1:n'] = [p] * (d - 1) + [g]$
        \EndIf
        \State $n' = n' - d$
    \EndIf
\EndWhile
\EndFunction
\end{algorithmic}
\caption{(Find $\mathbf{x}$)}
\end{algorithm}

\begin{algorithm}[H]
 \floatname{algorithm}{Indel Algorithm Step 2}
 \caption{}
\begin{algorithmic}[1]
\Function{DoStepTwo}{$\mathbf{x}$, $\mathbf{y}$, $d$, $q$, $m$, $n$, $n'$, $a$, $g$}
\State $b = d - a$
\State $p = q - 1$
\State $\mathbf{x}^{(1)} = [0] * (d-1) + [g + 1] + \mathbf{x}[n' + 1 : n]$
\State $\mathbf{x}^{(2)} = [p] * (d-1) + [g] + \mathbf{x}[n' + 1 : n]$
\State $v_1 = $ ComputeV($\mathbf{x}^{(1)}$, $y$, $n$, $n'$, $d$, $a$)
\State $v_2 = $ ComputeV($\mathbf{x}^{(2)}$, $y$, $n$, $n'$, $d$, $a$)
\If {$v_1 \geq v_2$}
    \State $t = 1$
\Else
    \State $t = 2$
\EndIf
\State $m'' = M(\mathbf{x}) - \sum_{i=n' - d + 1}^{n}x_{i-n'+d}^{(t)}w_i$

\If {$v_t \geq n -n' + 2a + 1$}
    \State \Return $t == 1$, $\Lambda$
\ElsIf{$v_t == n - n' + 2a$}
    \State $\mathbf{y'} = \mathbf{y}[1:len(\mathbf{y})-v_t-b]$
    \State possible\_result = DecodeDeletions($\mathbf{y'}$, $m''$, $d$, $n' - d$, $q$) + $\mathbf{x}^{t}$
    \If {AnswerIsCorrect(possible\_result, $M(\mathbf{x})$, $\mathbf{y}$, $n$, $b$)}
        \State \Return $\neg \text{$t == 1$}$, possible\_result
    \Else
        \State \Return \text{$t == 1$}, $\Lambda$
    \EndIf
\Else
    \For{$j = 1$ to $len(\mathbf{y}) - v_t - b$}
        \State $\mathbf{y'} = \mathbf{y}[1:j-1] + \mathbf{y}[j+1:len(\mathbf{y})-v_t-b]$
        \State possible\_result = DecodeDeletions($\mathbf{y'}$, $m''$, $d$, $n' - d$, $q$) + $\mathbf{x}^{t}$
        \If {AnswerIsCorrect(possible\_result, $M(\mathbf{x})$, $\mathbf{y}$, $n$, $b$)}
            \State \Return $\neg \text{$t == 1$}$, possible\_result
        \EndIf
    \EndFor
    \If{$len(\mathbf{y})-v_t-b \leq 0$} \Comment{If this is true, the above loop has not run}
        \State $\mathbf{y'} = \Lambda$
        \State possible\_result = DecodeDeletions($\mathbf{y'}$, $m''$, $d$, $n' - d $, $q$) + $\mathbf{x}^{t}$
        \If {AnswerIsCorrect(possible\_result, $M(\mathbf{x})$, $\mathbf{y}$, $n$, $b$)}
            \State \Return $\neg \text{$t == 1$}$, possible\_result
        \EndIf
    \EndIf
    \State \Return \text{$t == 1$}, $\Lambda$

\EndIf
\EndFunction
\Function{DecodeDeletions}{$\mathbf{y}$, m, d, n, q}
    \Comment{Implemented as described in \cite{deletion_algo_paper}}
\EndFunction

\Function{AnswerIsCorrect}{possible\_result, original\_moment, $\mathbf{y}$, $n$, $b$}
\State \Return $M(\text{possible\_result}) == $ original\_moment and $lcs(\text{possible\_result}, \mathbf{y}) \geq n - b$
\EndFunction
\end{algorithmic}
\caption{}
\end{algorithm}

\subsection{\bf Moment Algorithm} \label{algo:moment_algorithm_definition}

Given a corrupted codeword $\mathbf{y}$, and the values of $n$, $d$, and $r$, this algorithm finds the value of $M(\mathbf{x})$. \newline

Let $\mathbf{y}' = \mathbf{y}$. Let $i > 0$ be the lowest integer such that, for all $i \leq j < len(\mathbf{y}')$, we have $y'_{j} \geq y'_{j+1}$. Now, let $\mathbf{y}'$ be $\mathbf{y}$ with $y_i$ deleted. Repeat this process $a$ times. If $M(\mathbf{y}') \leq r$, then $M(\mathbf{x}) = r$. Else, $M(\mathbf{x}) = r + w_{n+1}$.

\begin{algorithm}[H]
 \floatname{algorithm}{Moment Algorithm}
 \caption{}
\begin{algorithmic}[1]
\Function{FindOriginalMoment}{$\mathbf{y}$, $n$, $d$, $r$}
\State $\mathbf{y}'=\mathbf{y}$\indent\Comment{Initialize $\mathbf{y}'$ to $\mathbf{y}$}
\For{$j = 1$ to $a$} \Comment{Do $a$ deletions}
    \State $i = len(\mathbf{y}')$
    \While{$i > 1$ and $y'_{i - 1} > y'_i$}
        \State $i = i - 1$
    \EndWhile
    \State $\mathbf{y}' = \mathbf{y}'[1:i-1] + \mathbf{y}'[i+1:len(\mathbf{y})]$ \Comment{Delete $y'_i$ from $\mathbf{y}'$}
\EndFor
\If{$M(\mathbf{y}') \leq r$}
    \State \Return $r$
\Else
    \State \Return $r + w_{n+1}$
\EndIf	
\EndFunction
\end{algorithmic}
\caption{(Find M(\textbf{x}))}
\end{algorithm}

\section{Proof of Indel-Correcting Algorithm} \label{Proof}

\subsection{Moment Theorem}

The following Moment Theorem allows us to recover the moment of any codeword $\mathbf{x}\in C_H(n,d,r,q)$ from its corruption $\mathbf{y}$ before decoding. Their proofs require the fact that $0 \leq M(\mathbf{x}) \leq \sum_{i=1}^{n}pw_i < 2w_n$ as shown in Lemma 10 in the Appendix of \cite{deletion_algo_paper}, and written here in \cref{lem:m_2_values}.

We present the Moment Theorem in two parts: Part 1 describes an algorithm for performing $a$ deletions on $\mathbf{y}$ (where $a$ is an upper bound on the number of insertions errors that occurred in $\mathbf{x}$) that minimizes the moment of the resulting codeword. Part 2 proves that the resulting codeword has moment less than or equal to $r$ if and only if $M(\mathbf{x}) \leq r$. The Moment Algorithm computes $M(\mathbf{x})$ based on these two parts.

For Part 1, some intuition can be obtained by analogy by considering the algorithm, given a number written in any base, to minimize the value of that number by deleting a fixed number of digits. Our algorithm is similar in spirit, including its proof.

\begin{theorem}[Moment Theorem - Part 1]\label{thm:moment_thm_p1}
    Given the corrupted codeword $\mathbf{y}$ and the ability to choose $\mathbf{\tilde{x}}$ as $\mathbf{y}$ with $a$ deletions (where $a$ is the upper bound on the number of insertions defined earlier), the following is a way to minimize the value of $M(\mathbf{\tilde{x}})$.

    Let ${\mathbf{y}'} = \mathbf{y}$. Let $i > 0$ be the lowest integer such that, for all $i \leq j < len(\mathbf{y}')$, we have $y'_{j} \geq y'_{j+1}$ (this means $i$ is the index of the left-most bit of the longest nonincreasing substring ending at the right-most character of $\mathbf{y}$). Now, let $\mathbf{y}'$ be $\mathbf{y}$ with $y_i$ deleted. Repeat this process $a$ times. Let $\mathbf{\tilde{x}}$ be the final value of $\mathbf{y}'$.
\end{theorem}
\begin{proof}
    Given a set of indices $S$, we will write $\mathbf{y} - S$ to represent $\mathbf{y}$ after deleting all the digits with indices in $S$. Let $i > 0$ be defined as in Theorem \ref{thm:moment_thm_p1}.
    We will prove that given a set of deleted indices $I = \{i_1, i_2, ...i_a\}$ that does not contain $i$, it is possible to replace one element of $I$ with $i$ without increasing the moment of $\mathbf{y}-I$.  It follows by induction that this algorithm is correct.

    For such a set $I$, at least one of the following must be true:
    
    \begin{itemize}
        \item There exists some $i_k \in I$ immediately to the right of $i$ (in other words, an $i_k \in I$ such that $i < i_k$ and there does not exist an $i_j \in I$ such that $i < i_j < i_k$).
        \item There exists some $i_k \in I$ immediately to the left of $i$ (in other words, an $i_k \in I$ such that $i_k < i$ and there does not exist an $i_j \in I$ such that $i_k < i_j < i$).
    \end{itemize}

    Define $I' = (I \cup \{i\}) \setminus \{i_k\}$.  To prove that $M(\mathbf{y} - I') \leq M(\mathbf{y} - I)$ for the first case, we calculate as follows:
    \begin{align}
        M(\mathbf{y} - I) - M(\mathbf{y} - I') & =
        \sum_{j=i}^{i_k - 1}y_jw_{j - k + 1} - \sum_{j=i+1}^{i_k}y_jw_{j - k} \\
        & = \sum_{j=i}^{i_k - 1}y_jw_{j - k + 1} - \sum_{j=i}^{i_k - 1}y_{j+1}w_{j - k + 1} \\
        & = \sum_{j=i}^{i_k - 1}(y_j - y_{j+1})w_{j - k + 1}\\
        & \geq 0
    \end{align}
    The last line is because $y_j \geq y_{j+1}$ for all $i \leq j < len(\mathbf{y})$, by our definition of $i$.

    As for the second case, we have
    \begin{align}
        M(\mathbf{y} - I) - M(\mathbf{y} - I') & = \sum_{j=i_k+1}^{i}y_jw_{j-k} - \sum_{j=i_k}^{i-1}y_jw_{j-k+1} \\
        & = y_iw_{i-k} + \sum_{j=i_k+1}^{i-1}y_j(w_{j-k} - w_{j-k+1}) - y_{i_k}w_{i_k -k + 1} \\
        & = y_iw_{i-k} + y_{i-1}(w_{i-k-1} - w_{i-k}) + \sum_{j=i_k+1}^{i-2}y_j(w_{j-k} - w_{j-k+1}) - y_{i_k}w_{i_k -k + 1} \label{moment_thm_note} \\
        & \geq y_iw_{i-k} + y_{i-1}(w_{i-k-1} - w_{i-k}) + \sum_{j=i_k+1}^{i-2}p(w_{j-k} - w_{j-k+1}) - pw_{i_k -k + 1} \\
        & \geq y_iw_{i-k} + y_{i-1}(w_{i-k-1} - w_{i-k}) + \sum_{j=i_k+1}^{i-2}pw_{j-k} -\sum_{j=i_k+1}^{i-2} pw_{j-k+1} - pw_{i_k -k + 1} \\
        & \geq y_iw_{i-k} + y_{i-1}(w_{i-k-1} - w_{i-k}) + \sum_{j=i_k+1}^{i-2}pw_{j-k} -\sum_{j=i_k}^{i-2} pw_{j-k+1} \\
        & \geq y_iw_{i-k} + y_{i-1}(w_{i-k-1} - w_{i-k}) + \sum_{j=i_k+1}^{i-2}pw_{j-k} -\sum_{j=i_k+1}^{i-1} pw_{j-k} \\
        & \geq y_iw_{i-k} + y_{i-1}(w_{i-k-1} - w_{i-k}) - pw_{i - k - 1} \\
        & \geq (y_{i-1} + 1)w_{i-k} + y_{i-1}(w_{i-k-1} - w_{i-k}) - pw_{i - k - 1} \label{moment_thm_note_2} \\
        & \geq w_{i-k} + (y_{i-1} - p)w_{i-k-1} \\
        & \geq 1 + p\sum_{j=i-k-d}^{i-k-1}w_j - pw_{i-k-1} \\
        & \geq 1 + p\sum_{j=i-k-d}^{i-k-2}w_j \\
        & > 0
    \end{align}

    We note that in \cref{moment_thm_note}, we assume that $i_k < i - 1$; otherwise, the proof follows easily from the previous line. Also, in \cref{moment_thm_note_2}, $y_i > y_{i-1}$ from our definition of $i$, which implies $y_i \geq y_{i-1} + 1$.

    We now employ induction to prove that the algorithm works for all values of $a$. Our base case, $a = 1$, follows from the above proof of $M(\mathbf{y} - I') \leq M(\mathbf{y} - I)$.  Next, assume that the algorithm is valid when $a = k$, i.e., given a string, we can perform the optimal $k$ deletions to minimize the value of the moment of the result.

    If $a = k + 1$, then the earlier part of the proof establishes that there exists an optimal set of indices to delete that contains $i$, which is the index we delete first. Then, because the algorithm works when $a = k$, it will be able to find the remainder of the indices to delete. So, the proof is complete.
\end{proof}

\begin{theorem}[Moment Theorem - Part 2] \label{thm:moment_thm_p2}
Suppose we are using a codebook where $M(\mathbf{x}) \mod {w_{n+1}}= r$.

Perform $a$ (recall that $a$ is an upper bound on the number of insertions that happened to $\mathbf{x}$) deletions to $\mathbf{y}$ to minimize the moment of the resulting codeword as described in \Cref{thm:moment_thm_p1}. Call the result of this $\tilde{\mathbf{x}}$. Then:

\begin{enumerate}
    \item If $M(\tilde{\mathbf{x}}) > r$, then $M(\mathbf{x}) = r + w_{n+1}$.
    \item If $M(\tilde{\mathbf{x}}) \leq r$, then $M(\mathbf{x}) = r$.
\end{enumerate}



\end{theorem}
\begin{proof}
For the first statement, we note that $\mathbf{y}$ is $\mathbf{x}$ after at most $b$ deletions and at most $a$ insertions. By performing at most $a$ deletions on $\mathbf{y}$, it is possible to undo all insertions. Let $\mathbf{\tilde{\tilde{x}}}$ be $\mathbf{y}$ after undoing all insertions. Since we created $\tilde{\mathbf{x}}$ by performing $a$ deletions to $\mathbf{y}$ to minimize the moment, and we created $\mathbf{\tilde{\tilde{x}}}$ by performing at most $a$ deletions on $\mathbf{y}$, we have $M(\mathbf{\tilde{x}}) \leq M(\mathbf{\tilde{\tilde{x}}})$. Also, since deletions can only decrease the value of the moment, and $\mathbf{\tilde{\tilde{x}}}$ can be made from $\mathbf{x}$ with only deletions, we have $ M(\mathbf{\tilde{\tilde{x}}}) \leq M(\mathbf{x})$. So, if $M(\mathbf{\tilde{x}}) > r$, we have $M(\mathbf{x}) > r$, which means $M(\mathbf{x}) = r + w_{n+1}$ since $r$ and $r + w_{n+1}$ are the only possible values of $M(\mathbf{x})$ (by \cref{lem:m_2_values}).

Now, we will show the second statement. Since $\mathbf{y}$ is $\mathbf{x}$ after at most $b$ deletions and at most $a$ insertions, this means $\tilde{\mathbf{x}}$ is $\mathbf{x}$ with at most $a + b = d$ deletions and at most $a$ insertions. Let $\tilde{\mathbf{x}}'$ be $\mathbf{x}$ with $d$ deletions
made as described in \Cref{thm:moment_thm_p1} to minimize the moment of $\tilde{\mathbf{x}}'$. Because insertions can only increase the moment of a codeword, we can see that $M(\tilde{\mathbf{x}}') \leq M(\tilde{\mathbf{x}})$. 

Let the indices of $\mathbf{x}$ that must be deleted to reach $\tilde{\mathbf{x}}'$ be labeled $i_1, i_2, ... i_{d}$ in increasing order. Intuitively, the following estimates use the fact that, when performing deletions, the largest possible difference between the starting and ending moments is obtained when the original codeword consists of all $p$'s. We now prove that $M(\tilde{\mathbf{x}}') \leq M(\tilde{\mathbf{x}})$:
\begin{align}
    M(\mathbf{x}) - M(\mathbf{\tilde{x}}') & = \sum_{j=1}^{n}x_jw_j - (\sum_{j=1}^{i_1 - 1}x_jw_j + \sum_{j=i_1+1}^{i_2 - 1}x_jw_{j-1} + ...) \\ 
    & = \sum_{j=1}^{n}x_jw_j - \sum_{k=1}^{d + 1} \sum_{j=i_{k-1} + 1}^{i_k - 1}x_jw_{j-k+1} \text{ (defining $i_0 = 0$ and $i_{d + 1} = n + 1$)}  \\
    & = \sum_{k=1}^{d}x_{i_k}w_{i_k} + \sum_{k=1}^{d + 1} \sum_{j=i_{k-1} + 1}^{i_k - 1}(x_jw_j - x_jw_{j-k+1}) \\
    & \leq \sum_{k=1}^{d}pw_{i_k} + \sum_{k=1}^{d + 1} \sum_{j=i_{k-1} + 1}^{i_k - 1}p(w_j - w_{j-k+1}) \\
    & \leq \sum_{j=1}^{n}pw_j - \sum_{k=1}^{d + 1} \sum_{j=i_{k-1} + 1}^{i_k - 1}pw_{j-k+1} \\
    & \leq \sum_{j=1}^{n}pw_j - \sum_{j=1}^{n - d} pw_{j} \\
    & \leq \sum_{j=n - d + 1}^{n}pw_j \\
    & < w_{n+1}
\end{align}

This implies $M(\mathbf{x}) < w_{n+1} + M(\tilde{\mathbf{x}}')$. We earlier had $M(\tilde{\mathbf{x}}') \leq M(\tilde{\mathbf{x}})$, so we now have $M(\mathbf{x}) < w_{n+1} + M(\tilde{\mathbf{x}})$. We now can see that if $M(\tilde{\mathbf{x}}) \leq r$, then $M(\mathbf{x}) < w_{n+1} + r$, so $M(\mathbf{x}) = r$ (again because $r$ and $r + w_{n+1}$ are the only possible values of $M(\mathbf{x})$ by \cref{lem:m_2_values}).

\end{proof}

\subsection{Proof of Step 1 of the Indel-Correcting Algorithm}

The following lemma shows the possible range of $m'$, given a value of $x_{n'}$. This is necessary to show the bounds in Step 1.

\begin{lemma}\label{lem:inrange}
If $x_{n'} = g$, then $gw_{n'} \leq m' \leq gw_{n'} + p\sum_{i=1}^{n' - 1} w_i$.
\end{lemma}

\begin{proof}
We have 
\begin{align}
m' & = M(\mathbf{x}) - \sum_{i=n' + 1}^{n} x_iw_i \\
& = \sum_{i=1}^{n}x_iw_i - \sum_{i=n' + 1}^{n} x_iw_i \\
& = \sum_{i=1}^{n'}x_iw_i \\
& = x_{n'}w_{n'} + \sum_{i=1}^{n' - 1}x_iw_i \\
& = gw_{n'} + \sum_{i=1}^{n' - 1}x_iw_i \label{lemma_proof:summation}
\end{align}

Since $0 \leq x_i \leq p$, we can bound the summation in \ref{lemma_proof:summation} by

\begin{equation}
0 \leq \sum_{i=1}^{n' - 1}x_iw_i \leq p\sum_{i=1}^{n' - 1}w_i
\end{equation}
and thus
\begin{equation}
gw_{n'} \leq m' \leq gw_{n'} + p\sum_{i=1}^{n' - 1}w_i
\end{equation}

\end{proof}

The following lemma is useful for making estimates.

\begin{lemma}\label{lem:n-2}
For all $c \geq 1$ and $d \geq 2$,
\begin{equation}
    w_c > p\sum_{i=1}^{c-2} w_i
\end{equation}
\end{lemma} 

\begin{proof}
We will prove this with induction. For the base case of $c = 1$, we have: 

\[w_1 = 1 > 0 = p\sum_{i=1}^{-1}w_i \] 

For the inductive case, assume it is true for some $c\geq 1$. We have
\begin{align}
w_{c + 1} & = 1 + p\sum_{i = c - d + 1}^{c} w_i\\
&= 1 + p\sum_{i = c - d + 1}^{c - 1} w_i + pw_c\\
&> 1 + p\sum_{i = c - d + 1}^{c - 1} w_i + p^2\sum_{i = 1}^{c - 2} w_i \text{ 
   (by our inductive assumption)}\\
&> pw_{c - 1} + 1 + p\sum_{i = 1}^{c - 2} w_i\\
&> p\sum_{i = 1}^{c - 1} w_i
\end{align}

This completes the proof.
\end{proof}

The following lemma shows that, for any $m'$, there are at most two possible values of $x_{n'}$, and they must be adjacent.

\begin{lemma}\label{lem:adjacent}
There are no integers $g_1$ and $g_2$ such that $\left|g_1 - g_2\right|  \geq 2$, $g_1w_{n'} \leq m' \leq g_1w_{n'} + p\sum_{i=1}^{n' - 1} w_i$, and $g_2w_{n'} \leq m' \leq g_2w_{n'} + p\sum_{i=1}^{n' - 1} w_i$.

\end{lemma}

\begin{proof}
Suppose for the sake of contradiction that such a $g_1$ and $g_2$ exist. Without loss of generality, take $g_1 < g_2$. We combine the above inequalities to get:
\begin{equation}
g_2 w_{n'} \leq m' \leq g_1w_{n'} + p\sum_{n=1}^{n'-1} w_i
\end{equation}

We subtract $g_1w_{n'}$ from both sides to obtain
\begin{equation}
(g_2 - g_1)w_{n'} \leq p\sum_{n=1}^{n'-1} w_i
\end{equation}
Since $g_2 - g_1 \geq 2$, we obtain
\begin{equation}
2w_{n'} \leq p\sum_{n=1}^{n'-1} w_i
\end{equation}

By the previous lemma, we have $w_{n'} > p\sum_{i=1}^{n' - 2} w_i$, which yields
\begin{equation}
w_{n'} + p\sum_{i=1}^{n' - 2} w_i< 2 w_{n'} \leq p\sum_{n=1}^{n'-1} w_i
\end{equation}
Finally, subtracting the summation on the left from both sides, we obtain
\begin{equation}
w_{n'} < pw_{n' - 1}
\end{equation}
This contradicts the recursive definition of the weights $w_{n'}$, so we are done.

\end{proof}

\subsection{Proof of Step 2 of the Indel-Correcting Algorithm}

The following lemma is used as a piece of the next theorem.
\begin{lemma}\label{lem:two_cases_building_block}
    Given $n'$ and $j$ such that $n ' - d < j < n'$, we have:
    \begin{equation}
    w_{n'} + w_j > p\sum_{i = 1}^{n' - 1}w_i
    \end{equation}
\end{lemma}
\begin{proof}

We have
\begin{align}
    w_{n'} + w_j & = w_j + 1 + p\sum_{i=n' -d}^{n' - 1} w_i \text{  (by the recurrence of $w_{n'}$)} \\
    & > p\sum_{i=1}^{j-2}w_i + 1 + p\sum_{i=n' -d}^{n' - 1} w_i \text{   (by \Cref{lem:n-2})} \\
    & > p\sum_{i=1}^{n' - 1}w_i 
\end{align}
The last estimate is because $n' - d < j \implies j - 2 \geq n' -d - 1$.  
\end{proof}

Using the information gained from knowing Step 1 failed to find $x_{n'}$, this theorem shows we can break the problem into two possible cases.

\begin{theorem}\label{thm:two_cases}

If, for some $g$, $gw_{n'} \leq m' \leq gw_{n'} + p\sum_{i=1}^{n' - 1} w_i$ and $(g + 1)w_{n'} \leq m' \leq (g + 1)w_{n'} + p\sum_{i=1}^{n' - 1} w_i$, then one of the following is true:

\begin{itemize}
\item Case 1: $x_{n'} = g + 1$ and $ x_{i} = 0$ for all $n' - d < i < n'$.
\item Case 2: $x_{n'} = g$ and $ x_{i} = p$ for all $n' - d < i < n'$. 
\end{itemize}
\end{theorem}
\noindent These are the same cases as defined in \ref{step:2} of the Indel-Correcting Algorithm, but notated differently.
\begin{proof}
The overall structure of this proof is to show the following three parts:

\begin{enumerate}
    \item Either $x_{n'} = g$ or $x_{n'} = g + 1$ \label{proof:item_1}
    \item $n' - d > 1$, which tells us that there are at least $d + 1$ characters before $x_{n'}$ in $\mathbf{x}$ \label{proof:item_2}
    \item If $x_{n'} = g$, then $x_i = p$ for all $n' - d < i < n'$, and if $x_{n'} = g + 1$, then $x_i = 0$ for all $n' - d < i < n'$ \label{proof:item_3}
\end{enumerate}

To show Part \ref{proof:item_1}, assume for the sake of contradiction that $x_{n'} \neq g$ and $x_{n'} \neq g + 1$. By \Cref{lem:inrange}, $x_{n'}w_{n'}, \leq m' \leq x_{n'}w_{n'} + p\sum_{i=1}^{n' - 1} w_i$. But, if $x_{n'} > g + 1$, then we get a contradiction through \Cref{lem:adjacent} by using $g$ as $g_1$ and $x_{n'}$ as $g_2$, and if $x_{n'} < g$, then we get a contradiction through the same lemma using $g + 1$ as $g_1$ and $x_{n'}$ as $g_2$.

Next, to show Part \ref{proof:item_2}, we have $gw_{n'} \leq m' \leq gw_{n'} + p\sum_{i=1}^{n' - 1} w_i$ and $(g + 1)w_{n'}\leq m' \leq (g + 1)w_{n'} + p\sum_{i=1}^{n' - 1} w_i$, so combining these inequalities we get:

\[gw_{n'} + p\sum_{i=1}^{n' - 1} w_i \geq m' \geq (g + 1)w_{n'}\]
Solving for the summation we obtain:
\[p\sum_{i=1}^{n' - 1} w_i \geq w_{n'}\]
Then, we use the recurrence of the weights:
\[p\sum_{i=1}^{n' - 1} w_i \geq 1 + p\sum_{i=n' - d}^{n' - 1} w_i\]

Comparing the starting index of the sum on the right-hand side, we must have $n' - d > 1$ for the inequality to be satisfied.

Now, we consider Part \ref{proof:item_3}, beginning with the case where $x_{n'} = g$. Assume for the sake of contradiction that there is some $n' - d < j < n'$ such that $x_{j} \neq p$. We have
\begin{equation}
    w_{n'} + w_j > p\sum_{i = 1}^{n'-1}w_i \text{   by \Cref{lem:two_cases_building_block}}
\end{equation}
Adding $gw_{n'}$ to both sides and rearranging:
\begin{equation}
    (g+1)w_{n'} > gw_{n'} - w_j + p\sum_{i = 1}^{n'-1}w_i
\end{equation}
Using $m' \geq (g + 1)w_{n'}$ (given as an assumption in the theorem statement), we obtain
\begin{equation}
    m' > gw_{n'} - w_j + p\sum_{i = 1}^{n'-1}w_i
\end{equation}

On the other hand, we have
\begin{align}
    m' & = M(\mathbf{x}) - \sum_{i=n' + 1}^{n} x_iw_i\\
        & = \sum_{i=1}^{n'} x_iw_i\\
        & = gw_{n'} + \sum_{i=1}^{n' - 1} x_iw_i  \text{   (because $x_{n'} = g$)} \\
        & \leq gw_{n'} - w_j + p\sum_{i=1}^{n' - 1} x_iw_i  \text{   (because $p$ is the maximum value of any $x_i$, and $x_j \neq p$)}
\end{align}
Thus, we have shown $m' > gw_{n'} - w_j + p\sum_{i = 1}^{n'-1}w_i$ and $m' \leq gw_{n'} - w_j + p\sum_{i = 1}^{n'-1}w_i$, which is a contradiction. 

Now, we will consider the second case where $x_{n'} = g+ 1$, and assume for the sake of contradiction that there is some $n' -d < j < n'$ such that $x_j \neq 0$. The following proof is very similar to what we just did. We have
\begin{equation}
    w_{n'} + w_j > p\sum_{i = 1}^{n'-1}w_i \text{   by \Cref{lem:two_cases_building_block}}
\end{equation}
Once again, we add $gw_{n'}$ to both sides:
\begin{equation}
    (g+1)w_{n'} + w_j > gw_{n'} + p\sum_{i = 1}^{n'-1}w_i
\end{equation}
We again use $m' \leq gw_{n'} + p\sum_{i=1}^{n'-1}w_i$ to obtain
\begin{equation}
    (g+1)w_{n'} + w_j > m'
\end{equation}

On the other hand, we have
\begin{align}
    m' & = M(\mathbf{x}) - \sum_{i=n' + 1}^{n} x_iw_i\\
        & = \sum_{i=1}^{n'} x_iw_i\\
        & = (g+1)w_{n'} + \sum_{i=1}^{n' - 1} x_iw_i  \text{   (because $x_{n'} = g + 1$)} \\
        & \geq gw_{n'} + w_j \text{   (because $0$ is the minimum value of any $x_i$, and $x_j \neq 0$)}
\end{align}
Thus, we have $m' \geq gw_{n'} + w_j$ and $m' < gw_{n'} + w_j$, which is once again a contradiction. The proof is complete.

\end{proof}

The following lemma establishes a lower bound for $v$. This is necessary to have a polynomial time complexity (with a constant exponent, as opposed to one that depends on $d$).

\begin{lemma}\label{thm:v_lower_bound}

If we consider

\item $\mathbf{x}[n' - d + 1 : n] = \mathbf{x}^{(1)} = [0] * (d - 1) + [g + 1] + x[n' + 1 : n]$ (Case 1) 
\item $\mathbf{x}[n' - d + 1 : n] = \mathbf{x}^{(2)} = [p] * (d - 1) + [g] + x[n' + 1 : n]$ (Case 2)

Then $\max(v_1, v_2) \geq n - n' + 2a - 1$, where $v_1$ and $v_2$ are defined as they were in \textbf{Step 2 \ref{def:v_definition}} of the Indel-Correcting Algorithm. 

\end{lemma}

\begin{proof}

If $\max(v_1, v_2) = \infty$, we are done, so suppose that $\max(v_1, v_2) < \infty$. 
We define $\mathbf{y}_{end} = \mathbf{y}[len(\mathbf{y}) - max(v_1, v_2) + 1 : len(\mathbf{y})]$.

First, we see that:
\begin{align}
    lcs(\mathbf{x}^{(1)}, \mathbf{x}^{(2)}) & = lcs([0] * (d - 1) + [g + 1] + x[n' + 1 : n], [p] * (d - 1) + [g] + x[n' + 1 : n]) \\
    & \leq lcs([0] * (d-1), [p] * (d-1)) + (n - n' + 1) \\
    & \leq n - n' + 1
\end{align}

By \Cref{lem:lcs}, for any strings $s_1, s_2, s_3$, we have $len(s_1) \geq lcs(s_1, s_2) + lcs(s_1, s_3) - lcs(s_2, s_3)$. Therefore:
\begin{align}
    len(\mathbf{y}_{end}) & \geq lcs(\mathbf{y}_{end}, \mathbf{x}^{(1)}) + lcs(\mathbf{y}_{end}, \mathbf{x}^{(2)}) - lcs(\mathbf{x}^{(1)}, \mathbf{x}^{(2)}) \\
    & \geq lcs(\mathbf{x}^{(1)}, \mathbf{y}[len(\mathbf{y}) - v_1 + 1 : len(\mathbf{y})]) + lcs(\mathbf{x}^{(2)}, \mathbf{y}[len(\mathbf{y}) - v_2 + 1 : len(\mathbf{y})]) - (n - n' + 1) \\
    & \geq (n - n' + d - b) + (n - n' + d - b) - (n -n' + 1) \\
    & \geq n - n' + 2(d-b) - 1 \\
    & \geq n - n' + 2a - 1
\end{align}

Since $len(\mathbf{y}_{end}) = max(v_1, v_2)$, this proves the lemma.

\end{proof}

We are guaranteed by \Cref{lem:lcs} to have a $j$ where $lcs(\mathbf{x}, \mathbf{y}) = lcs(\mathbf{x}[1 : n' - d], \mathbf{y}[1 : j]) + lcs(\mathbf{x}[n' - d + 1 : n], \mathbf{y}[j + 1: len(\mathbf{y})])$. This lemma rules out possible values of $j$, which is necessary in the lemmas to follow.

Also, we should note that many of the following lemmas assume Case $t$ is correct (as a reminder, $t$ is the index such that $\max(v_1, v_2) = v_t$). This implies $\mathbf{x}[n' - d + 1 : n] = \mathbf{x}^{(t)}$. This fact is used frequently.

\begin{lemma} \label{lem:j_upper_bound}
    If Case $t$ is correct, then there is no $j$ such that $0 \leq j \leq len(\mathbf{y})$, and ($j > len(\mathbf{y}) - v_t$ or $j < n' - d - b$), where $lcs(\mathbf{x}, \mathbf{y}) = lcs(\mathbf{x}[1 : n' - d], \mathbf{y}[1 : j]) + lcs(\mathbf{x}[n' - d + 1 : n], \mathbf{y}[j + 1: len(\mathbf{y})])$.
\end{lemma}

\begin{proof}

We know $lcs(\mathbf{x}, \mathbf{y}) \geq n - b$ by \Cref{lem:n_b}. Now, assume for the sake of contradiction that a $j > len(\mathbf{y}) - v_t$ exists. It follows that
\begin{align}
    n - b & \leq lcs(\mathbf{x}, \mathbf{y}) \\ 
    & \leq lcs(\mathbf{x}[1 : n' - d], \mathbf{y}[1 : j]) + lcs(\mathbf{x}[n' - d + 1 : n], \mathbf{y}[j + 1: len(\mathbf{y})]) \\
    & \leq n' - d + lcs(\mathbf{x}[n' - d + 1 : n], \mathbf{y}[j + 1: len(\mathbf{y})]) \\
    & \leq n' - d + lcs(\mathbf{x}[n' - d + 1 : n], \mathbf{y}[len(\mathbf{y}) - v_t + 2: len(\mathbf{y})]) \text{ 
   because $j > len(\mathbf{y}) - v_t$} \\
   & \leq n' - d + lcs(\mathbf{x}[n' - d + 1 : n], \mathbf{y}[len(\mathbf{y}) - (v_t - 1) + 1: len(\mathbf{y})]) \\
    & < n' - d + (n - n' + d - b) \text{   by definition of $v_t$ and that Case $t$ is correct} \\
    & < n - b
\end{align}
So, we have reached a contradiction. If instead $j < n' - d - b$, we have
\begin{align}
    n - b & \leq lcs(\mathbf{x}, \mathbf{y}) \\ 
    & \leq lcs(\mathbf{x}[1 : n' - d], \mathbf{y}[1 : j]) + lcs(\mathbf{x}[n' - d + 1 : n], \mathbf{y}[j + 1: len(\mathbf{y})]) \\
    & \leq j + (n - (n' - d + 1) + 1) \\
    & \leq j + n - n' + d \\
    & < n' - d - b + n - n' + d \\
    & < n - b
\end{align}
This again is a contradiction, so the lemma is proven.
\end{proof}

This lemma is used in later lemmas - in an informal sense, it is used to show that $\mathbf{y}[1:len(\mathbf{y}) - v - b]$ is similar enough to $\mathbf{x}[1 : n' - d]$ for the Deletions Algorithm to go from the former (or a small modification of the former) to the latter.

\begin{lemma}\label{thm:matching_first_part_of_y}
    If Case $t$ is correct, then $lcs(\mathbf{y}[1:len(\mathbf{y}) - v_t - b], \mathbf{x}[1 : n' - d]) \geq n' - d - b$.
\end{lemma}

\begin{proof}

    By \Cref{lem:lcs}, there exists some $0 \leq j \leq len(\mathbf{y})$ such that $lcs(\mathbf{x}, \mathbf{y}) = lcs(\mathbf{x}[1 : n' - d], \mathbf{y}[1 : j]) + lcs(\mathbf{x}[n' - d + 1 : n], \mathbf{y}[j + 1: len(\mathbf{y})])$. Consider the lowest value of $j$ that satisfies this equation. Note that in this proof, since we are assuming Case $t$ is correct, $\mathbf{x}^{(t)} = \mathbf{x}[n' - d + 1 : n]$, but for easier manipulation $\mathbf{x}^{(t)}$ is not used in this proof. Also, more importantly, note that $lcs(\mathbf{x}, \mathbf{y}) \geq n - b$ by \Cref{lem:n_b}.

    By \Cref{lem:j_upper_bound}, $n' - d - b \leq j \leq len(\mathbf{y}) - v_t$. Also, by  \Cref{thm:v_lower_bound}, $v_t \geq n - n' + 2a - 1$.

Now, assume for the sake of contradiction that $lcs(\mathbf{y}[1:len(\mathbf{y}) - v_t - b], \mathbf{x}[1 : n' - d]) < n' - d - b$. We consider two cases. First, assume $j < len(\mathbf{y}) - v_t - b$. It follows that
\begin{align}
    n - b & \leq lcs(\mathbf{x}, \mathbf{y}) \\ 
    & \le lcs(\mathbf{x}[1 : n' - d], \mathbf{y}[1 : j]) + lcs(\mathbf{x}[n' - d + 1 : n], \mathbf{y}[j + 1: len(\mathbf{y})]) \\ 
    & \leq lcs(\mathbf{x}[1 : n' - d], \mathbf{y}[1 : len(\mathbf{y}) - v_t - b]) + lcs(\mathbf{x}[n' - d + 1 : n], \mathbf{y}[j + 1: len(\mathbf{y})]) \label{eq:j_less_than_case} \\ 
    & < n' - d - b + lcs(\mathbf{x}[n' - d + 1 : n], \mathbf{y}[j + 1: len(\mathbf{y})]) \\ 
    & < n' - d - b + n - n' + d \\ 
    & < n - b
\end{align}
which is a contradiction. We note that in \cref{eq:j_less_than_case}, we used the fact that $j < len(\mathbf{y}) - v_t - b$. We now consider the other case, where $j \geq len(\mathbf{y}) - v_t - b$.  Similarly, we find that
\begin{align}
    n - b & \leq lcs(\mathbf{x}, \mathbf{y}) \\ 
    & \leq lcs(\mathbf{x}[1 : n' - d], \mathbf{y}[1 : j]) + lcs(\mathbf{x}[n' - d + 1 : n], \mathbf{y}[j + 1: len(\mathbf{y})]) \\ 
    & \leq lcs(\mathbf{x}[1 : n' - d], \mathbf{y}[1 : len(\mathbf{y}) - v_t - b]) + (j - (len(\mathbf{y}) - v_t - b)) \label{eq:j_greater_than} \\
    & \quad  + lcs(\mathbf{x}[n' - d + 1 : n], \mathbf{y}[j + 1: len(\mathbf{y})]) \\ 
    & < n' - d - b + (j - (len(\mathbf{y}) - v_t - b)) + lcs(\mathbf{x}[n' - d + 1 : n], \mathbf{y}[j + 1: len(\mathbf{y})]) \label{eq:j_less_than_upper} \\
    & < n' - d - b + (j - (len(\mathbf{y}) - v_t - b)) \\ 
    & \quad + lcs(\mathbf{x}[n' - d + 1 : n], \mathbf{y}[len(\mathbf{y}) - v_t + 1: len(\mathbf{y})]) + (len(\mathbf{y}) - v_t - j) \\
    & < n' - d - b + (j - (len(\mathbf{y}) - v_t - b)) + (n - n' + d - b) + (len(\mathbf{y}) - v_t - j) \text{  (by \Cref{lem:v_definition_equivs})} \\
    & < n' - d - b + j - (len(\mathbf{y}) - v_t - b) + n - n' + d - b + len(\mathbf{y}) - v_t - j \\
    & < n - b
\end{align}
yielding a contradiction.  Thus, the theorem is proven. We note that in \cref{eq:j_greater_than,eq:j_less_than_upper}, we used the fact that $j \geq len(\mathbf{y}) - v_t - b$ and $j \leq len(\mathbf{y}) - v_t$, respectively.
\end{proof}


The next lemma proves \textbf{Step 2 \ref{algo:step_2_best_case}} of the Indel-Correcting Algorithm.
\begin{lemma}\label{thm:v_upper_bound}
    If $v_t \geq n - n' + 2a + 1$, Case $t$ is incorrect.
\end{lemma}

\begin{proof}

Assume to the contrary that Case $t$ is correct.

If we have $v_t = \infty$: There is no value of $v$ such that $lcs(\mathbf{x}[n'-d+1:n], \mathbf{y}[len(\mathbf{y})-v+1:len(\mathbf{y})]) = n - n' + d - b$. Therefore, we have $lcs(\mathbf{x}[n'-d+1:n], \mathbf{y}) < n - n' + d - b$. By \Cref{lem:lcs}, this implies  $lcs(\mathbf{x}, \mathbf{y}) < n - n' + d - b + n' - d = n - b$. However, we know $lcs(\mathbf{x}, \mathbf{y}) \geq n -b$ by \Cref{lem:n_b}, so this is a contradiction, so this case must be incorrect.

If we do not have $v_t = \infty$: By \Cref{thm:matching_first_part_of_y}, we have $lcs(\mathbf{y}[1:len(\mathbf{y}) - v_t - b], \mathbf{x}[1 : n' - d]) \geq n' - d - b$. We see:
\begin{align}
    n' - d - b & \leq lcs(\mathbf{y}[1:len(\mathbf{y}) - v_t - b], \mathbf{x}[1 : n' - d]) \\
    & \leq len(\mathbf{y}) - v_t - b \\
    & \leq n + a - 2b - v_t \\
    & \leq n + a - 2b - (n - n' + 2a + 1)\\
    & \leq n' - a - 2b - 1\\
    & \leq n' - d - b - 1
\end{align}
So, since we showed $n' - d - b \leq n' - d - b - 1$, we have reached a contradiction and the theorem is proven.
\end{proof}




The following lemma proves \textbf{Step 2 \ref{algo:step_2_middle_case}} of the Indel-Correcting Algorithm.
\begin{lemma}\label{thm:v_is_n_n_prime_2a}
    
    If Case $t$ is correct, and $v_t = n - n' + 2a$, then the following process is guaranteed to find $\mathbf{x}$:

    \item Let $m'' = M(\mathbf{x}) - \sum_{i=n' - d + 1}^{n}x_{i-n'+d}^{(t)}w_i$. Use the Deletions Algorithm for the codebook $C_H(n' - d, d, m'', q)$ to decode $\mathbf{y'} \equiv \mathbf{y}[1 : len(\mathbf{y}) - v_t - b]$. Call the result of decoding $\mathbf{x'}$.

    $\mathbf{x}$ will be $\mathbf{x'} + \mathbf{x}^{(t)}$ (in the actual algorithm, we check the correctness of the solution because we do not have the first hypothesis of the theorem - we do not know if Case $t$ is correct).   
\end{lemma}

\begin{proof}

We see that $m'' = M(\mathbf{x}[1 : n' - d])$.Therefore, we know $\mathbf{x}[1 : n' - d] \in C_H(n' - d, d, m'', q)$. 

From \Cref{thm:matching_first_part_of_y}, we know $lcs(\mathbf{y}[1:len(\mathbf{y}) - v_t - b], \mathbf{x}[1 : n' - d]) \geq n' - d - b$, and we also see:
\begin{align}
    len(\mathbf{y}) - v_t - b & = (n + a - b) - (n - n' + 2a) - b \\
    & = - b + n' - a - b \\
    & = n' - d - b
\end{align}

So, it follows that $\mathbf{y'} = \mathbf{y}[1:len(\mathbf{y}) - v_t - b]$ can be obtained from $\mathbf{x}[1 : n' - d]$ with $b$ deletions, which means that the Deletion Algorithm for $C_H(n' -d , d, m'', q)$ will decode $\mathbf{x}[1 : n' - d]$ when given $\mathbf{y'}$. Since we assumed we are working in the correct case, we have $\mathbf{x}^{(t)} = \mathbf{x}[n' - d + 1:n]$, so (as we are calling the output of the deletions algorithm $\mathbf{x'}$), we have $\mathbf{x} = \mathbf{x'} + \mathbf{x}^{(t)}$.
\end{proof}

The next lemma proves \textbf{Step 2 \ref{algo:step_2_worst_case}} of the Indel-Correcting Algorithm.
\begin{lemma}\label{thm:v_is_n_n_prime_2a_minus_1}
If Case $t$ is correct, and $v_t = n - n' + 2a - 1$ in this case, then the following is guaranteed to find a solution:

\noindent If $len(\mathbf{y}) - v_t - b > 0$:
\begin{itemize}
\item
For each $1 \leq j \leq len(\mathbf{y}) - v_t - b$, use the Deletions Algorithm for the codebook $C_H(n' - d, d, m'', q)$ to decode $\mathbf{y'} \equiv \mathbf{y}[1:j-1] + \mathbf{y}[j + 1 : len(\mathbf{y}) - v_t - b]$ for up to $d$ deletions. Call the result of decoding $\mathbf{x'}$.

For each $j$, check if $M(\mathbf{x'} + \mathbf{x}^{(t)}) = M(\mathbf{x})$ and if $\mathbf{x'} + \mathbf{x}^{(t)}$ shares a subsequence of at least $n - b$ with $\mathbf{y}$. This must be true for at least one $\mathbf{x'}$; for this $\mathbf{x'}$ we have $\mathbf{x} = \mathbf{x'} + \mathbf{x}^{(t)}$.
\end{itemize}
\noindent If $len(\mathbf{y}) - v_t - b \leq 0$:
\begin{itemize}
\item
Use the Deletions Algorithm for the codebook $C_H(n' -d, d, m'', q)$ to decode $\mathbf{y'} \equiv \Lambda$. Call the result $\mathbf{x'}$.
The solution will be $\mathbf{x} = \mathbf{x'} + \mathbf{x}^{(t)}$.

\end{itemize}

\end{lemma}

\begin{proof}

We see that $m'' = M(\mathbf{x}[1 : n' - d])$.Therefore, we know $\mathbf{x}[1 : n' - d] \in C_H(n' - d, d, m'', q)$. 

From \Cref{thm:matching_first_part_of_y}, we know $lcs(\mathbf{y}[1:len(\mathbf{y}) - v_t - b], \mathbf{x}[1 : n' - d]) \geq n' - d - b$, and we also see:
\begin{align}
    len(\mathbf{y}) - v_t - b & = (n + a - b) - (n - n' + 2a - 1) - b \\
    & = - b + n' - a - b + 1 \\
    & = n' - d - b + 1
\end{align}

We will break the proof into cases:

\begin{enumerate}
        \item $len(\mathbf{y}) - v_t - b \leq 0$: By the above, this implies $n' - d - b + 1 \leq 0$, so $n' - d \leq b - 1 < d$. Therefore, $\Lambda$ is $\mathbf{x}[1:n' - d]$ after less than $d$ deletions, so the Deletions Algorithm for $C_H(n' -d, d, m'', q)$ will find $\mathbf{x}[1:n' - d]$ successfully.
        \item $len(\mathbf{y}) - v_t - b > 0$ and $lcs(\mathbf{y}[1:len(\mathbf{y}) - v_t - b], \mathbf{x}[1 : n' - d]) = n' - d - b + 1$: Then $\mathbf{y}[1 : len(\mathbf{y}) - v_t - b]$ can be obtained from $\mathbf{x}[1 : n' - d]$ with $b - 1$ deletions, so $\mathbf{y}[1 : len(\mathbf{y}) - v_t - b - 1]$ can be obtained from $\mathbf{x}[1 : n' - d]$ with $b$ deletions. This means the Deletions Algorithm for $C_H(n' -d, d, m'', q)$ will find $\mathbf{x}[1:n' - d]$ when $j = len(\mathbf{y}) - v_t - b$, since that means $\mathbf{y'} = \mathbf{y}[1 : j - 1] + \mathbf{y}[j + 1 : len(\mathbf{y}) - v_t - b] = \mathbf{y}[1 : len(\mathbf{y}) - v_t - b - 1]$.
        \item $len(\mathbf{y}) - v_t - b > 0$ and $lcs(\mathbf{y}[1:len(\mathbf{y}) - v_t - b], \mathbf{x}[1 : n' - d]) = n' - d - b$:
        In this case, by \Cref{lem:lcs}, we know there exists some $1 \leq j \leq len(\mathbf{y}) - v_t - b$ such that $lcs(\mathbf{y}[1:j-1] + \mathbf{y}[j + 1 : len(\mathbf{y}) - v_t - b], \mathbf{x}[n' - d]) = lcs(\mathbf{y}[1:len(\mathbf{y}) - v_t - b], \mathbf{x}[n' - d])$. For such a $j$, we see that $len(\mathbf{y}[1:j-1] + \mathbf{y}[j + 1 : len(\mathbf{y}) - v_t - b]) =n' - d - b = lcs(\mathbf{y}[1:j-1] + \mathbf{y}[j + 1 : len(\mathbf{y}) - v_t - b], \mathbf{x}[n' - d])$, so $\mathbf{y}[1:j-1] + \mathbf{y}[j + 1 : len(\mathbf{y}) - v_t - b]$ can be obtained from $\mathbf{x}[1:n'-d]$ with $b$ deletions, so the Deletions Algorithm for $C_H(n' -d, d, m'', q)$ will be able to find $\mathbf{x}[1:n'-d]$ with this $j$.
\end{enumerate}

As in the previous lemma, since we assumed we are working in the correct case, we have $\mathbf{x}^{(t)} = \mathbf{x}[n' - d + 1:n]$, so when $\mathbf{x'} = \mathbf{x}[1:n'-d]$, we have $\mathbf{x} = \mathbf{x'} + \mathbf{x}^{(t)}$.  We note the last two cases rely on \Cref{lem:check_if_correct} to guarantee that an incorrect answer cannot be derived.
\end{proof}

\section{Decoding Examples} \label{Examples}
We provide examples of decoding using the Indel-Correcting Algorithm.

\begin{example}
Consider the following example:

\begin{itemize}
    \item $n = 10$, $d = 3$, $r = 381$, $q = 2$
    \item $\mathbf{x} = 001\textcolor{red}{1}110001$ 
    \item $\mathbf{y} = 001110001\textcolor{blue}{01}$ 
\end{itemize}

A red (blue) digit indicates that the corresponding bit in $\mathbf{x}$ was deleted (inserted, respectively).

We can verify that $M(\mathbf{x}) \equiv 381 \mod w_{n+1}$, by generating the weights and calculating the moment:

\begin{table}[H]
\centering
\begin{tabular}{|c|c|c|c|c|c|c|c|c|c|c|c|}
\hline
\textbf{$w_0$} & \textbf{$w_1$} & \textbf{$w_2$} & \textbf{$w_3$} & \textbf{$w_4$} & \textbf{$w_5$} & \textbf{$w_6$} & \textbf{$w_7$} & \textbf{$w_8$} & \textbf{$w_9$} & \textbf{$w_{10}$} & \textbf{$w_{11}$} \\
\hline
0 & 1 & 2 & 4 & 8 & 15 & 28 & 52 & 96 & 177 & 326 & 600 \\
\hline
\end{tabular}
\vspace{1em}
\caption{Weights with $d = 3$, $q = 2$}
\end{table}
\vspace{-2em}

We see $M(\mathbf{x}) = 0 * 1 + 0 * 2 + 1 * 4 + 1 * 8 + 1 * 15 + 1 * 28 + 0 * 52 + 0 * 96 + 0 * 177 + 1 * 326 = 381$, and $381 \mod 600 = 381$ (this check is not part of the algorithm - it is just here to demonstrate an example of how moments are computed).

\begin{itemize}

\item Perform Preliminary \ref{algo:a_b_d_fix} We calculate $a = \lfloor\frac{d + len(\mathbf{y}) - n}{2}\rfloor = \lfloor\frac{3 + 11 - 10}{2}\rfloor = 2$ and $b = \lfloor\frac{3 - 11 + 10}{2}\rfloor = 1$. Since $a + b = d$, we do not perform an extra deletion on $\mathbf{y}$. 

\item Perform Preliminary \ref{algo:moment_algorithm_usage} We now use the Moment Algorithm to find $M(\mathbf{x})$. Since $a =2$, we will perform 2 deletions on $\mathbf{y}$. We see the lowest $i > 0$ such that $\mathbf{y}_j \geq \mathbf{y}_{j+1}$ for all $i \leq j < len(\mathbf{y})$ is $i = 11$. So, we delete $y_{11}$, and get $\mathbf{y}' = 0011100010$. The next deletion is with $i = 9$, so we get $\mathbf{y}' = 001110000$. Computing $M(\mathbf{y}') = 27 \leq r = 381$, we see by the Moment Theorem that $M(\mathbf{x}) = 381$. We now apply the Indel Algorithm.

\item We now set $n' = n = 10$, and do \ref{step:1}. We calculate $m' = M(\mathbf{x}) - \sum_{i=11}^{10}x_iw_i = M(\mathbf{x}) = 381$. We then find $H = \{0, 1\}$, because $0 \leq m' \leq \sum_{i=1}^{n'-1}w_i = 383$, and $w_{n'} = 326 \leq m' \leq w_{n'} + \sum_{i=1}^{n'-1}w_i = 709$. So, we enter \ref{step:2}.

\item We see $\mathbf{x}^{(1)} = 001$ and $\mathbf{x}^{(2)} = 110$. We compute $v_1 = 2$ and $v_2 = 3$. Because $v_2 > v_1$, we have $t = 2$. We see $v_2 = n - n' + 2a - 1 = 3$, so we enter \textbf{Step 2 \ref{algo:step_2_worst_case}}.

\item We compute $m'' = M(\mathbf{x}) - \sum_{i=n'-d+1}^{n}x_{i-n'+d}^{(2)}w_i = 108$. We see $len(\mathbf{y}) - v_t - b = 11 - 3 - 1 = 7$, so for each $1 \leq j \leq 7$, we use the Deletions Algorithm to decode $\mathbf{y}[1:j-1] + \mathbf{y}[j+1:len(\mathbf{y}) - v_t - b]$. The Deletions Algorithm is not the focus of this paper, so this process will not be shown here.

The $\mathbf{x'}$ returned from the Deletions Algorithm for each $j$ are $0111001$, $0111001$, $0011001$, $0011001$, $0011001$, $0011101$, and $0011101$.

We see that, for each of these, $M(\mathbf{x'} + \mathbf{x}^{(t)}) \neq 381$, so we rule out $\mathbf{x}^{(2)}$, so $\mathbf{x}^{(1)}$ must be correct. Therefore, $x_{10} = 1$, $x_9 = 0$, and $x_8 = 0$. We now re-enter \ref{step:1}, set $n' = 7$, and calculate $m' = M(\mathbf{x}) - \sum_{i=8}^{10}x_iw_i = 55$. Once again, $H = \{0, 1\}$, because $0 \leq m' \leq \sum_{1}^{6} w_i = 59$, and $w_7 = 52 \leq m' \leq w_7 + \sum_{1}^{6} w_i  = 111$. So, we enter step 2 again.

\item We see $\mathbf{x}^{(1)} = 001001$ and $\mathbf{x}^{(2)} = 110001$. We compute $v_1 = 5$ and $v_2 = 7$. Because $v_2 > v_1$, we have $t = 2$. We see $v_2 = n - n' + 2a = 7$, so we enter \textbf{Step 2 \ref{algo:step_2_middle_case}}. 

\item We compute $m'' = M(\mathbf{x}) - \sum_{i=n'-d+1}^{n}x_{i-n'+d}^{(2)}w_i = 12$. We see $len(\mathbf{y}) - v_t - b = 11 - 7 - 1 = 3$. We use the Deletions Algorithm on $\mathbf{y}[1:len(\mathbf{y}) - v_t - b] = \mathbf{y}[1:3] = 001$. We use the Deletions Algorithm to decode this, and get the result $\mathbf{x}' = 0011$. We test $M(\mathbf{x}' + \mathbf{x}^{(2)}) = M(0011110001) = 381 = M(\mathbf{x})$, and that $lcs(\mathbf{x}' + \mathbf{x}^{(2)}, y) \geq n - b = 9$, so we are done, and $\mathbf{x} = \mathbf{x}' + \mathbf{x}^{(2)} = 0011110001$.

\end{itemize}
\end{example}

\begin{example}
We will now look at an example where we never enter step 2. Here, we have:

\begin{itemize}
    \item $n = 10$, $d = 3$, $r = 434$, $q = 3$
    \item $\mathbf{x} = 10212102\textcolor{red}{2}2$
    \item $\mathbf{y} = 10212102\textcolor{blue}{0}2$
\end{itemize}

The weights table with $d = 3, q = 3$, for reference:

\begin{table}[H]
\centering
\begin{tabular}{|c|c|c|c|c|c|c|c|c|c|c|c|}
\hline
\textbf{$w_0$} & \textbf{$w_1$} & \textbf{$w_2$} & \textbf{$w_3$} & \textbf{$w_4$} & \textbf{$w_5$} & \textbf{$w_6$} & \textbf{$w_7$} & \textbf{$w_8$} & \textbf{$w_9$} & \textbf{$w_{10}$} & \textbf{$w_{11}$} \\
\hline
0 & 1 & 3 & 9 & 27 & 79 & 231 & 675 & 1971 & 5755 & 16803 & 49059 \\
\hline
\end{tabular}
\vspace{1em}
\caption{Weights with $d = 3$, $q = 3$}
\end{table}
\vspace{-2em}
\begin{itemize}
\item As in the previous example, we can verify $\mathbf{x} \in C_H(n, d, r, q)$ by observing $M(\mathbf{x}) = 1 * 1 + 3 * 0 + 9 * 2 + 27 * 1 + 79 * 2 + 231 * 1 + 675 * 0 + 1971 * 2 + 5755 * 2 + 16803 * 2 = 49493$, and $49493 \mod 49059 = 434$.

\item Perform Preliminary \ref{algo:a_b_d_fix} We calculate $a = \lfloor\frac{d + len(\mathbf{y}) - n}{2}\rfloor = \lfloor\frac{3 + 10 - 10}{2}\rfloor = 1$ and $b = \lfloor\frac{3 - 10 + 10}{2}\rfloor = 1$. Since $a + b < d$, we perform an extra deletion on $\mathbf{y}$. So, we run the algorithm with $\mathbf{y} = 021210202$. Now, $a = \lfloor\frac{d + len(\mathbf{y}) - n}{2}\rfloor = \lfloor\frac{3 + 9 - 10}{2}\rfloor = 1$ and $b = \lfloor\frac{3 - 9 + 10}{2}\rfloor = 2$.

\item Perform Preliminary \ref{algo:moment_algorithm_usage} We use the Moment Algorithm to find the value of $M(\mathbf{x})$. Since $a = 1$, we perform one deletion on $\mathbf{y}$. We see the lowest $i > 0$ such that $y_j \geq y_{j+1}$ for all $i \leq j < len(\mathbf{y})$ is $i = 9$. So, we delete $y_9$, and get $\mathbf{y}' = 02121020$. Computing $M(\mathbf{y}') = 1498 > 434 = r$. Therefore, $M(\mathbf{x}) = r + w_{n+1} = 434 + 49059 = 49493$.

\end{itemize}

Since the next example shows the process of \ref{step:1} in detail, and this example only includes step 1, here we will only give a table of $n', m'$, and $x_{n'}$:

\begin{table}[H]
\centering
\begin{tabular}{|c|c|c|}
\hline
\textbf{$n'$} & \textbf{$m'$} & \textbf{$x_{n'}$} \\
\hline
10 & 49493 & 2 \\
9  & 15887 & 2 \\
8  & 4377  & 2 \\
7  & 435   & 0 \\
6  & 435   & 1 \\
5  & 204   & 2 \\
4  & 46    & 1 \\
3  & 19    & 2 \\
2  & 1     & 0 \\
1  & 1     & 1 \\
\hline
\end{tabular}
\vspace{1em}

\caption{Values of $n'$, $m'$, and $x_{n'}$ in Example 3}
\end{table}

\vspace{-2em}

We have decoded $x = 1021210222$.
\end{example}

\begin{example}

Now, we will look at an example with a non-binary alphabet:

\begin{itemize}
    \item $n = 9$, $d = 2$, $r = 147376$, $q = 4$
    \item $\mathbf{x} = 130\textcolor{red}{2}00103$ 
    \item $\mathbf{y} = \textcolor{blue}{0}13002103$
    
\end{itemize}

The weights table with $d = 2, q = 4$, for reference:

\begin{table}[H]
\centering
\begin{tabular}{|c|c|c|c|c|c|c|c|c|c|c|c|}
\hline
\textbf{$w_0$} & \textbf{$w_1$} & \textbf{$w_2$} & \textbf{$w_3$} & \textbf{$w_4$} & \textbf{$w_5$} & \textbf{$w_6$} & \textbf{$w_7$} & \textbf{$w_8$} & \textbf{$w_9$} & \textbf{$w_{10}$} \\
\hline
0 & 1 & 4 & 16 & 61 & 232 & 880 & 3337 & 12652 & 47968 & 181861 \\
\hline
\end{tabular}
\vspace{1em}
\caption{Weights with $d = 2$, $q = 4$}

\end{table}
\vspace{-2em}

\begin{itemize}

\item As in the previous examples, we can verify $\mathbf{x} \in C_{H}(n, d, r, q)$ by observing $M(\mathbf{x}) = 1 * 1 + 4 * 3 + 16 * 0 + 61 * 2 + 232 * 0 + 880 * 0 + 3337 * 1 + 12652 * 0 + 47968 * 3 = 147376$, and $147376 \mod w_{10} = 147376 \mod 181861 = 147376$.

\item Perform Preliminary \ref{algo:a_b_d_fix} We calculate $a = \lfloor\frac{d + len(\mathbf{y}) - n}{2}\rfloor = \lfloor\frac{2 + 9 - 9}{2}\rfloor = 1$ and $b = \lfloor\frac{2 - 9 + 9}{2}\rfloor = 1$. Since $a + b = d$, we do not perform an extra deletion on $\mathbf{y}$.

\item Perform Preliminary \ref{algo:moment_algorithm_usage} We use the Moment Algorithm to find $M(\mathbf{x})$. Since $a = 1$, we will perform 1 deletion on $\mathbf{y}$. We see the lowest $i > 0$ such that $\mathbf{y}_j \geq \mathbf{y}_{j+1}$ for all $i \leq j < len(\mathbf{y})$ is $i = 9$. So, we delete $y_9$, and get $\mathbf{y}' = {0}1300210$. Computing $M(\mathbf{y}') = 5149 \leq r = 147376$, we see $M(\mathbf{x}) = 147376$.

\item We now set $n' = n = 9$, and do \ref{step:1}. We calculate $m' = M(\mathbf{x}) = \sum_{10}^{9}x_iw_i = 147376$. We see $H = \{2, 3\}$, since $2w_9 \leq m' \leq 2w_9 + 3\sum_{i=1}^{8}w_i$, and $3w_9 \leq m' \leq 3w_9 + 3\sum_{i=1}^{8}w_i$. So, we enter \ref{step:2}.

\item We see $\mathbf{x}^{(1)} = 03$ and $\mathbf{x}^{(2)} = 32$. We compute $v_1, v_2 = 1$. We can choose either value of $t$ since $v_1 = v_2$; we will choose $t = 2$. We see $v_t = n - n' + 2a - 1$, so we enter \textbf{Step 2 \ref{algo:step_2_worst_case}}.

\item We compute $m'' = M(\mathbf{x}) - \sum_{i=8}^{9}\mathbf{x}^{(2)}_iw_i = 13484$, and $len(\mathbf{y}) - v_t - b = 7$. So, for each $1 \leq j \leq 7$, we use the Deletions Algorithm to decode $\mathbf{y}[1:j-1] + \mathbf{y}[j + 1 : len(\mathbf{y}) - v_t - b]$. As in the first example, the process of the Deletions Algorithm will be skipped.

The results from the Deletions Algorithm for each $j$ are $1300013$, $0300013$, $0100013$, $0130013$, $0130013$, $0130013$, and $0130003$.

\item We see that, for each of these, $M(\mathbf{x'} + \mathbf{x}^{(2)}) \neq M(\mathbf{x})$, so we rule out $\mathbf{x}^{(2)}$. Therefore, $x_9 = 3$ and $x_8 = 0$. We now set $n' = 7$, and calculate $m' = M(\mathbf{x}) - \sum_{i=8}^{9}x_iw_i = 3472$. Now, we see $H = \{0, 1\}$, since $0 \leq m' \leq 3\sum_{i=1}^{6}w_i$ and $w_7 \leq m' \leq w_7 + 3\sum_{i=1}^{6}w_i$. So, we enter \ref{step:2} again.

\item We see $\mathbf{x}^{(1)} = 0103$ and $\mathbf{x}^{(2)} = 3003$. We see $v_1 = 3$ and $v_2 = 4$, so we let $t = 2$. Since $v_t = n - n' + 2a = 4$, we enter \textbf{Step 2 \ref{algo:step_2_middle_case}}. We see $m'' = M(\mathbf{x}) - \sum_{i=6}^{9}x_i^2w_i = 832$. We put $\mathbf{y}[1:len(\mathbf{y}) - v_t - b] = \mathbf{y}[1:9 - 4 - 1] = \mathbf{y}[1:4] = 0130$ into the Deletions Algorithm. We get $\mathbf{x'} = 01303$ as a result, but $M(\mathbf{x'} + \mathbf{x}^{(2)}) = M(013033003) \neq M(\mathbf{x})$, so we know $\mathbf{x}^{(2)}$ is incorrect. Therefore, $x_7 = 1$ and $x_6 = 0$. 

\item We do step \ref{step:1} again, now with $n' = 5$, and calculate $m' = 135$. This time, $H = \{0\}$, since $0 \leq m' \leq 3\sum_{i=1}^{4}w_i$. Since $|H| = 1$, $x_5$ is the only element in $H$, so $x_5 = 0$. 

\item We go to step \ref{step:1} again, setting $n' = 4$. We compute $m' = 135$, and find $H = \{2\}$, since $2w_4 \leq m' \leq 2w_4 + 3\sum_{i=1}^{3}w_i$. Since $|H| = 1$, $x_4$ is the only element in $H$, so $x_4 = 2$.

\end{itemize}

The remaining three bits of $x$ go similarly; the values of $n', m'$, and $x_{n'}$ are given in the table below:

\begin{table}[H]
\centering
\begin{tabular}{|c|c|c|}
\hline
\textbf{$n'$} & \textbf{$m'$} & \textbf{$x_{n'}$} \\
\hline
3 & 13 & 0 \\
2 & 13 & 3 \\
1 & 1  & 1 \\
\hline
\end{tabular}
\vspace{1em}
\caption{Remaining values}
\end{table}

We are now done, having found $\mathbf{x} = 130200103$.
\end{example}

\section{Appendix} \label{Appendix}

\begin{lemma}\label{lem:lcs}
    The following are useful lemmas about subsequences:
    \begin{itemize}
        \item For any strings $s, s_1, s_1$, we have $len(s) \geq lcs(s, s_1) + lcs(s, s_2) - lcs(s_1, s_2)$.
        \item For any strings $s_1$ and $s_2$, and any $0 \leq i \leq len(s_1)$, there exists $0 \leq j \leq len(s_2)$ such that $lcs(s_1, s_2) = lcs(s_1[1 : i], s_2[1: j]) + lcs(s_1[i+1:len(s_1)], s_2[j+1:len(s_2)])$.
        \item For any strings $s_1$ and $s_2$, and $j \geq 0$, we have $lcs(s_1, s_2) \leq j + lcs(s_1[1:len(s_1) - j], s_2[1:len(s_2) - j])$ and $lcs(s_1, s_2) \leq j + lcs(s_1[j+1:len(s_1)], s_2[j+1:len(s_2)])$.
        \item For any strings $s_1$ and $s_2$, and $j \geq 0$, we have $lcs(s_1, s_2) \leq j + lcs(s_1[1:len(s_1) - j], s_2)$ and $lcs(s_1, s_2) \leq j + lcs(s_1[j+1:len(s_1)], s_2)$.
        \item For any strings $s_1, s_2$, if $lcs(s_1, s_2) < len(s_1)$, there exists some $1 \leq j \leq len(s_1)$ such that $lcs(s_1, s_2) = lcs(s_1[1:j-1] + s_1[j + 1:len(s_1)], s_2)$.
        \item For any strings $s_1, s_2$, $len(s_1) \geq lcs(s_1, s_2)$. 
        \item For any strings $s_1, s_2$, if $s_1$ is not a subsequence of $s_2$, then $lcs(s_1, s_2) < len(s_1)$. 
    \end{itemize}

\end{lemma}

\begin{proof}
    First, we will prove that, for any strings $s, s_1, s_2$, we have $len(s) \geq lcs(s, s_1) + lcs(s, s_2) - lcs(s_1, s_2)$.

    Consider a longest common subsequence between $s$ and $s_1$ - call this $c_1$. Let $I^{(1)}$ be a set containing the indices of characters of $c_1$ within $s_1$ (so, if $I^{(1)}_k$ is the $k$th element of $I^{(1)}$ when ordered ascendingly, $c_1 = s_{I^{(1)}_1}s_{I^{(1)}_2}s_{I^{(1)}_3}...s_{I^{(1)}_{lcs(s, s_1)}}$). Similarly, define $c_2$ and $I^{(2)}$.

    For convenience, let $S = \{i \in \mathbb{Z} | 1 \leq i \leq len(s)\}$.
    We see:
    \begin{align}
        len(s) & = |S| \\
        & = |S \cap I^{(1)}| + |S \setminus I^{(1)}| \\
        & = |I^{(1)}| + |S \setminus I^{(1)}| \\
        & = |I^{(1)}| + |(S \setminus I^{(1)}) \cap I^{(2)}| + |(S \setminus I^{(1)}) \setminus I^{(2)}| \\
        & \geq |I^{(1)}| + |(S \setminus I^{(1)}) \cap I^{(2)}| \\
        & \geq |I^{(1)}| + |(I^{(2)} \setminus I^{(1)})| \\
        & \geq |I^{(1)}| + |(I^{(2)} \setminus I^{(1)})| + |(I^{(2)} \cap I^{(1)})| - |(I^{(2)} \cap I^{(1)})|
        & \geq |I^{(1)}| + |(I^{(2)}| - |(I^{(2)} \cap I^{(1)})| \\
        & \geq lcs(s, s_1) + lcs(s, s_2) - |(I^{(2)} \cap I^{(1)})|
    \end{align}

    Now, notice that $(I^{(2)} \cap I^{(1)})$ gives indices of a subsequence between $s_1$ and $s_2$, so $|(I^{(2)} \cap I^{(1)})| \leq lcs(s_1, s_2)$. Combined with the earlier lines, we have $len(s) \geq lcs(s, s_1) + lcs(s, s_2) - lcs(s_1, s_2)$ as desired.
\end{proof}
\begin{proof}
    Next, we will prove that for any strings $s_1$ and $s_2$, and any $0 \leq i \leq len(s_1)$, there exists $0 \leq j \leq len(s_2)$ such that $lcs(s_1, s_2) = lcs(s_1[1 : i], s_2[1:j]) + lcs(s_1[i+1:len(s_1)], s_2[j+1:len(s_2)])$.

    Consider a longest common subsequence $c$ between $s_1$ and $s_2$. Define $I^{(1)}$ to be a set containing the indices of characters of $c$ within $s_1$. Analogously, define $I^{(2)}$. Take any $0 \leq i \leq len(s_1)$. Let $q = |\{k \in I^{(1)} | k \leq i\}|$. Then, take $j$ to be the $q$th lowest integer in $I^{(2)}$, or $0$ if $q = 0$. 
    \newline We see that $\{k \in I^{(1)} | k \leq i\}$ and $\{k \in I^{(1)} | k \leq j\}$ define a common subsequence between $s_1[1:i]$ and $s_2[1:j]$, and likewise, $\{k \in I^{(1)} | k > i\}$ and $\{k \in I^{(1)} | k > j\}$ define a common subsequence between $s_1[i+1:len(s_1)]$ and $s_2[j+1:len(s_2)]$. Therefore:
    \begin{align}
        lcs(s_1, s_2) & = |I^{(1)}| \\
                      & = |\{k \in I^{(1)} | k \leq i\}| + |\{k \in I^{(1)} | k > i\}| \\
                      & \leq lcs(s_1[1 : i], s_2[1:j]) + lcs(s_1[i+1:len(s_1)], s_2[j+1:len(s_2)])
    \end{align}

    To show the inequality in the other direction, we see that if $c_1$ is a longest common subsequence between $s_1[1:i]$ and $s_2[1:j]$, and $c_2$ is a longest common subsequence between $s_1[i+1:len(s_1)]$ and $s_2[j+1:len(s_2)]$, then $c_1 + c_2$ is a common subsequence between $s_1$ and $s_2$, so:
    \begin{align}
        lcs(s_1, s_2) & \geq len(c_1 + c_2) \\ & \geq lcs(s_1[1 : i], s_2[1:j]) + lcs(s_1[i+1:len(s_1)], s_2[j+1:len(s_2)])
    \end{align}
\end{proof}
\begin{proof}
    Now, we will prove that for any strings $s_1$ and $s_2$, and $j \geq 0$, we have $lcs(s_1, s_2) \leq j + lcs(s_1[1:len(s_1) - j], s_2[1:len(s_2) - j])$ and $lcs(s_1, s_2) \leq j + lcs(s_1[j+1:len(s_1)], s_2[j+1:len(s_2)])$.

    The result listed below (that any strings $s_1$ and $s_2$, and $j \geq 0$, we have $lcs(s_1, s_2) \leq j + lcs(s_1[1:len(s_1) - j], s_2)$ and $lcs(s_1, s_2) \leq j + lcs(s_1[j+1:len(s_1)], s_2)$) clearly follows from the above.
    \newline \newline
    In the following proof, we will use the well-known fact that, for two non-empty strings $s_1$ and $s_2$, if ${s_{1,len(s_1)}} ={s_{2,len(s_2)}}$, then $lcs(s_1, s_2) = 1 + lcs(s_1[1:len(s_1) - 1], s_2[1: len(s_2) -1])$, and if ${s_{1,len(s_1)}} \neq{s_{2,len(s_2)}}$, $lcs(s_1, s_2) = max(lcs(s_1[1:len(s_1) - 1], s_2[1: len(s_2) ]), lcs(s_1[1:len(s_1) ], s_2[1: len(s_2) -1]))$.
    \newline \newline
    We will do this by induction. Our base case will be when $j + len(s_1) + len(s_2) = 0$, when we can see the inequality is clearly true.

    Now, we do the inductive case: Suppose this is true when $j + len(s_1) + len(s_2) \leq k$, for some $k$. Take $j, s_1, s_2$ such that $j + len(s_1) + len(s_2) = k + 1$. If $s_1$ or $s_2$ is the empty string, then $lcs(s_1, s_2) = 0$, and the inequality is true. Also, if $j = 0$, the inequality is trivially true. So, we assume ${s_{1,len(s_1)}}$ and  ${s_{2,len(s_2)}}$ exist, and $j \neq 0$. If ${s_{1,len(s_1)}} ={s_{2,len(s_2)}}$, then:

    \begin{align}
        lcs(s_1, s_2) & = 1 + lcs(s_1[1:len(s_1) - 1], s_2[1: len(s_2) -1]) \\ 
        &  \leq j + lcs(s_1[1:len(s_1) - j], s_2[1: len(s_2) -j]) \text{ (by our inductve hypothesis)}
    \end{align}

    If ${s_{1,len(s_1)}} \neq {s_{2,len(s_2)}}$, then:

    \begin{align}
        lcs(s_1, s_2) & = max(lcs(s_1[1:len(s_1) - 1], s_2[1: len(s_2)]), lcs(s_1[1:len(s_1) ], s_2[1: len(s_2) -1])) \\
        & \leq max(j + lcs(s_1[1:len(s_1) - (j + 1)], s_2[1: len(s_2) -j]), \\
        & \quad j + lcs(s_1[1:len(s_1) -j ], s_2[1: len(s_2) - (j+1)])) \text{   (by our inductive hypothesis)} \\
        & \leq j + lcs(s_1[1:len(s_1) - j], s_2[1: len(s_2) -j]
    \end{align}

    So, the proof is complete.
\end{proof}

The remaining statements are clear enough that they do not need a proof (the fifth statement becomes clear when noting that $s_1[1:j-1] + s_1[j+1:len(s_1)]$ is $s_1$ with the $j$th character deleted).

\begin{lemma}\label{lem:a_b_bounds}
    $a$ is an upper bound on the number of insertions, and $b$ is an upper bound on the number of deletions.
\end{lemma}

\begin{proof}
    Suppose there were $i$ insertions and $j$ deletions, and $i > a$. Then, $j \leq d - i$, so we have:

    \begin{align}
        len(\mathbf{y}) & = len(x) + i - j \\
        & = n + i - j \\
        & = n + a - j \\
        & \geq n + a - (d - i) \\
        & \geq n + a - b - a + i \\
        & \geq len(\mathbf{y}) - a + i \\
        & > len(\mathbf{y})
    \end{align}

    We have a contradiction, so $a$ is an upper bound for the number of insertions. The case for where $j > b$ is similar.
\end{proof}

\begin{lemma}\label{lem:n_b}
$lcs(\mathbf{x}, \mathbf{y}) \geq n - b$.
\end{lemma}
\begin{proof}
We see every character of $\mathbf{x}$ that is not deleted forms a common subsequence between $\mathbf{x}$ and $\mathbf{y}$, and we proved in \Cref{lem:a_b_bounds} that $b$ is an upper bound for the number of deletions. Therefore, $lcs(\mathbf{x}, \mathbf{y}) \geq len(\mathbf{x}) - b = n - b$. 
\end{proof}

\begin{lemma}[Lemma 10 of \cite{deletion_algo_paper}]\label{lem:m_2_values}
$0 \leq M(\mathbf{x}) \leq \sum_{i=1}^{n}pw_i < 2w_n$, which implies that for a given Helberg codebook and codeword $\mathbf{x}$, there are only 2 possible values of $M(\mathbf{x})$.
\end{lemma}







\begin{lemma}\label{lem:v_definition_equivs}
An equivalent definition for $v_i$ is:

For $i \in \{1, 2\}$, the lowest nonnegative integer $v_i$ such that $lcs(\mathbf{x}^{(i)}, \mathbf{y}[len(\mathbf{y}) - v_i + 1 : len(\mathbf{y})]) = n - n' + d - b$.
\end{lemma}
\begin{proof}
Define $v_i$ as normal (as the lowest nonnegative integer such that $lcs(\mathbf{x}^{(i)}, \mathbf{y}[len(\mathbf{y}) - v_i + 1 : len(\mathbf{y})]) \geq n - n' + d - b$), and define $z_i$ as the lowest nonnegative integer such that $lcs(\mathbf{x}^{(i)}, \mathbf{y}[len(\mathbf{y}) - z_i + 1 : len(\mathbf{y})]) = n - n' + d - b$. Because the condition for $z_i$ is stricter, we see that $z_i \geq v_i$. 

Now, suppose for the sake of contradiction that $z_i > v_i$. Then, $lcs(\mathbf{x}^{(i)}, \mathbf{y}[len(\mathbf{y}) - v_i + 1 : len(\mathbf{y})]) > n - n' + d - b$. We see this means $v_i$ is strictly positive, because $n - n + d - b \geq 0$, so $lcs(\mathbf{x}^{(i)}, \mathbf{y}[len(\mathbf{y}) - v_i + 1 : len(\mathbf{y})]) > n - n' + d - b$, and if $v_i = 0$ then $\mathbf{y}[len(\mathbf{y}) - v_i + 1 : len(\mathbf{y})]$ is the empty string.

By \Cref{lem:lcs}, we have $lcs(\mathbf{x}^{(i)}, \mathbf{y}[len(\mathbf{y}) - v_i + 2 : len(\mathbf{y})]) > n - n' + d - b - 1$, which implies $lcs(\mathbf{x}^{(i)}, \mathbf{y}[len(\mathbf{y}) - (v_i - 1) + 1 : len(\mathbf{y})]) \geq n - n' + d - b$. However, this means that $v_i - 1$ is a nonnegative integer lower than $v_i$ which satisfies $lcs(\mathbf{x}^{(i)}, \mathbf{y}[len(\mathbf{y}) - (v_i - 1) + 1 : len(\mathbf{y})]) \geq n - n' + d - b$, so $v_i - 1$ which is a contradiction. We are done. 
\end{proof}

\begin{lemma}\label{lem:check_if_correct}
    For any string $s$:
    
    $s = \mathbf{x}$ if and only if $M(s) = M(\mathbf{x})$, $lcs(s, \mathbf{y}) \geq n - b$, and $len(s) = n$.
\end{lemma}
\begin{proof}

First, we see that if $s = \mathbf{x}$, then $M(s) = M(\mathbf{x})$, $lcs(s, \mathbf{y}) = lcs(\mathbf{x}, \mathbf{y}) \geq n - b$ (by \Cref{lem:n_b}), and $len(s) = len(\mathbf{x}) = n$.

Now, we will show the other direction. By deleting all characters of $s$ that are not part of the longest common subsequence between $\mathbf{y}$ and $s$, and then inserting all characters in $\mathbf{y}$ that are not part of this longest common subsequence, we see that $\mathbf{y}$ is $s$ after at most $b$ deletions and at most $len(\mathbf{y}) - (n - b) = (n + a - b) - (n - b) = a$ insertions. Since $a + b = d$, we see that $s$ can be obtained from $\mathbf{y}$ after $d$ indels, which is also true of $\mathbf{x}$. Because $len(s) = n$ and $M(s) = M(\mathbf{x})$, $s$ is in the same Helberg codebook as $\mathbf{x}$, so together this implies that $s = \mathbf{x}$.
\end{proof}

\section{Acknowledgment}
The authors would like to thank Liam Busch, Jonathan Moore, Kaitlyn Myers, Emily Sandlin, and Amanda Swankoski for their early work on Helberg codes which provided key insights and preliminary results that led to the indel-correcting algorithm described in this paper.





\printbibliography

\end{document}